\documentclass[a4paper,12pt]{article}
\pdfoutput=1
\usepackage{graphicx, rotating,amssymb,footnote,cite,floatpag}
\makesavenoteenv{tabular}
\makesavenoteenv{table}

\ifx\pdfoutput\undefined
\usepackage[dvips,bookmarks]{hyperref}	
\else
\usepackage{hyperref}	
\fi
\hypersetup{colorlinks,bookmarksopen,bookmarksnumbered,citecolor=verdec,
linkcolor=blue,pdfstartview=FitH,urlcolor=rossoc}
\def\myurl#1#2{\href{http://#1}{#2}}
\def\hhref#1{\href{http://arxiv.org/abs/#1}{#1}} 

\usepackage{multicol}
\usepackage{color}
\definecolor{rosso}{cmyk}{0,1,1,0.4}
\definecolor{rossos}{cmyk}{0,1,1,0.55}
\definecolor{rossoc}{cmyk}{0,1,1,0.2}
\definecolor{blue}{cmyk}{1.0,0.20,0.3,0.1}
\definecolor{blus}{cmyk}{1,1,0,0.6}
\definecolor{bluc}{cmyk}{1,1,0,0.1}
\definecolor{verde}{cmyk}{0.92,0,0.59,0.25}
\definecolor{verdec}{cmyk}{0.92,0,0.70,0.05}
\definecolor{verdes}{cmyk}{0.92,0,0.59,0.4}

\font\tenrsfs=rsfs10 at 12pt
\font\sevenrsfs=rsfs7
\font\fiversfs=rsfs5
\newfam\rsfsfam
\textfont\rsfsfam=\tenrsfs
\scriptfont\rsfsfam=\sevenrsfs
\scriptscriptfont\rsfsfam=\fiversfs
\def\mathscr#1{{\fam\rsfsfam\relax#1}}

\def\circa#1{\,\raise.3ex\hbox{$#1$\kern-.75em\lower1ex\hbox{$\sim$}}\,}

\newcommand{\beq}{\begin{equation}}
\newcommand{\eeq}{\end{equation}}

\def\circa#1{\,\raise.3ex\hbox{$#1$\kern-.75em\lower1ex\hbox{$\sim$}}\,}
\makeatletter

\def\art{\@ifnextchar[{\eart}{\oart}}
\def\eart[#1]#2#3#4#5#6{{\rm #2}, {#3 #4} {\rm (#6) #5} [{\hhref{#1}}]}
\def\hepart[#1]#2{{\rm #2, \hhref{#1}}}
\newcommand{\oart}[5]{{\rm #1}, {#2 #3} {\rm (#5) #4}}

\newcounter{alphaequation}[equation]
\def\thealphaequation{\theequation\hbox to
0.6em{\hfil\alph{alphaequation}\hfil}}
\def\eqnsystem#1{
\def\@eqnnum{{\rm (\thealphaequation)}}
\def\@@eqncr{\let\@tempa\relax \ifcase\@eqcnt \def\@tempa{& & &} \or
  \def\@tempa{& &}\or \def\@tempa{&}\fi\@tempa
  \if@eqnsw\@eqnnum\refstepcounter{alphaequation}\fi
\global\@eqnswtrue\global\@eqcnt=0\cr}
\refstepcounter{equation} \let\@currentlabel\theequation \def\@tempb{#1}
\ifx\@tempb\empty\else\label{#1}\fi
\refstepcounter{alphaequation}
\let\@currentlabel\thealphaequation
\global\@eqnswtrue\global\@eqcnt=0 \tabskip\@centering\let\\=\@eqncr
$$\halign to \displaywidth\bgroup \@eqnsel\hskip\@centering
$\displaystyle\tabskip\z@{##}$&\global\@eqcnt\@ne
\hskip2\arraycolsep\hfil${##}$\hfil& \global\@eqcnt\tw@\hskip2\arraycolsep
$\displaystyle\tabskip\z@{##}$\hfil
\tabskip\@centering&\llap{##}\tabskip\z@\cr}
\def\endeqnsystem{\@@eqncr\egroup$$\global\@ignoretrue} \makeatother

\begin{document}
\begin{flushright}
{\footnotesize
{\sc IFT-UAM/CSIC-16-035}\\
}
\end{flushright}
\color{black}

\begin{center}
{\Huge\bf Updated galactic radio constraints\\[0.3cm]
on Dark Matter}

\medskip
\bigskip\color{black}\vspace{0.4cm}

{
{\large\bf Marco Cirelli}\ $^a$,
{\large\bf Marco Taoso}\ $^b$
}
\\[6mm]
{\it $^a$ \href{http://www.lpthe.jussieu.fr/spip/index.php}{Laboratoire de Physique Th\'eorique et Hautes Energies (LPTHE)},\\ UMR 7589 CNRS \& UPMC,\\ 4 Place Jussieu, F-75252, Paris, France}\\[3mm]
{\it $^b$ \href{https://www.ift.uam-csic.es/en}{Instituto de F\'isica Te\'orica (IFT)} UAM/CSIC,\\ calle Nicol\'as Cabrera 13-15, 28049 Cantoblanco, Madrid, Spain}
\end{center}

\begin{quote}
\color{black}\large

{\large\bf Abstract:} We perform a detailed analysis of the synchrotron signals produced by dark matter annihilations and decays. We consider different set-ups for the propagation of electrons and positrons, the galactic magnetic field and dark matter properties. We then confront these signals with radio and microwave maps, including {\sc Planck} measurements, from a frequency of 22 MHz up to 70 GHz.
We derive two sets of constraints: {\em conservative} and {\em progressive}, the latter based on a modeling of the astrophysical emission. Radio and microwave constraints are complementary to those obtained with other indirect detection methods, especially for dark matter annihilating into leptonic channels.
\end{quote}

\tableofcontents

\newpage

\section{Introduction}
\label{sec:introduction}

The existence of Dark Matter (DM) is well established via a number of cosmological and astrophysical probes (galactic rotation curves, lensing measurements, the Cosmic Microwave Background, Large Scale Structure formation), all pertaining to its gravitational effects. Other, more direct, manifestations of DM have been sorely lacking so far, despite the tremendous experimental and theoretical efforts, and are therefore eagerly sought after. 

A promising strategy is the one of Indirect Detection (ID): in general terms, it aims at identifying anomalous emissions which could be due to the annihilations or decays of DM particles, in our Galaxy or beyond. Such `anomalous emissions' are to be found in charged cosmic rays (electrons, positrons, antiprotons, antinuclei), neutrinos, gamma rays, X-rays or radio waves.
Despite several hints in the past in most of these channels, there is currently no clear indication of anomalies which can not be explained by astrophysics, even if sometimes of peculiar sort (see e.g.~\cite{Cirelli:2015gux} for a recent overview). In addition, and perhaps more importantly, there is no clear indication of a channel which could, by itself, provide an unambiguous signal. This is because, again, astrophysics is in general tightly intertwined with the anomalous features being looked for, making the task of disentangling them very challenging. In other words, a lesson learned in the latest few years in this field is that any convincing detection will have to be multi-messenger. 

\bigskip 

It is therefore important to diversify and multiply the ID lines of attack. In this paper we focus in particular on the radio-wave signals of galactic DM. Radio waves are produced by DM as secondary radiation, in the sense that they are emitted (via the synchrotron process) by the high-energetic electrons and positrons produced by DM annihilations or decays, as they propagate in the galactic magnetic field. This kind of signal has been studied and exploited many times in the past. 
Historically, people started by focussing on small regions around the Galactic Center (GC), since the expectation of a large DM density and the presence of a strong magnetic field together guarantee a significant emission there~\cite{Berezinsky:1992mx,Berezinsky:1994wva,Gondolo:2000pn,Bertone:2001jv,Bertone:2002je,Boehm:2002yz,Aloisio:2004hy,Hooper:2008zg,Regis:2008ij,Bertone:2008xr,Bergstrom:2008ag,Bringmann:2009ca,Crocker:2010gy,Boehm:2010kg,Linden:2011au,Laha:2012fg,Asano:2012zv,Bringmann:2014lpa,Cholis:2014fja}. However, uncertainties are also largest in that regions, both concerning the DM distribution and the astrophysical environment. 
More recently, a few works have shifted the focus to much larger regions (or the whole) of the galactic halo, which can provide more robust results~\cite{Blasi:2002ct,Borriello:2008gy,Regis:2009md,Delahaye:2011jn,Fornengo:2011iq,Mambrini:2012ue,Egorov:2015eta}. 
Outer galaxies and galaxy clusters have also been considered~\cite{Boehm:2002yz,Tasitsiomi:2003vw,Colafrancesco:2005ji,Colafrancesco:2006he,Siffert:2010cc,Boehm:2010kg,Carlson:2012qc,Regis:2014tga} as well as isotropic signals of extragalactic origin~\cite{Fornengo:2011cn,Fornengo:2011xk,Hooper:2012jc}.

\smallskip 

We are interested in re-examining the galactic halo analysis. We will be mostly based on the study in~\cite{Fornengo:2011iq}, and we aim at improving it in several respects. Namely, i) we use refined recent results~\cite{Buch:2015iya} for the production and propagation of electrons and positrons from DM, taking into account new effects both at low energy (e.g. losses due to bremsstrahlung and ionization) and at high energy (electroweak corrections affecting the DM annihilation/decay spectra); ii) we perform a systematic study considering a large range of DM masses, annihilation/decay channels, DM distribution profiles and $e^\pm$ propagation parameters; iii) we consider for the first time systematically the case of decaying DM~\footnote{Synchrotron emission from decaying DM has been considered in the past, e.g. in~\cite{Boehm:2010kg,Fornengo:2011cn}, but systematic bounds on the half-life have not been derived, to the best of our knowledge.}; iv) we use new observational maps at high frequency, namely those released by the {\sc Planck} satellite, which prove relevant for some channels and some ranges of DM masses (see below); v) we consider a modelization of the astrophysical galactic background.

\smallskip 

Our main outputs consist of fig.~\ref{fig:allbounds}, in which we present the constraints on DM annihilations or decays for a variety of primary channels. We derive two sets of constraints: {\em conservative constraints} (computed by just asking that the DM radio emission does not exceed the measured intensities, without any assumption on the astrophysical galactic contribution) and {\em progressive constraints} (computed by modeling the astrophysical emission and then adding the DM one).

\medskip

The rest of this paper is organized as follows. In sec.~\ref{sec:maps} we present the observational radio maps that we employ. In sec.~\ref{sec:formalism} we briefly review the main features of the radio emission from DM, with special attention to the differences between the different channels and the dependance on the astrophysical parameters. In sec.~\ref{sec:background} we present our modelization of the astrophysical emission. In sec.~\ref{sec:constraints} we specify the Regions of Interest (RoI) that we will use in the analysis and we present our results. In sec.~\ref{sec:conclusions} we provide our conclusions.


\section{Radio and microwave maps}
\label{sec:maps}

In this section we describe in some detail the radio and microwave emission galactic maps that we use. 

In our choices, the guiding principle is that we want to cover a wide range in frequencies, because (as explained below) different frequencies constraint different DM channels and DM masses. Frequencies above several tens of GHz are not interesting for our purpuses, since the DM signal typically peaks at lower frequencies, for DM masses below $\mathcal{O}$(10 TeV), and the galactic foreground (namely dust) starts to be overwhelming above $\sim 100$ GHz.
We also prefer maps that have an extensive coverage of the sky. 

\begin{figure}[t]
\begin{center}
\includegraphics[width= 0.33 \textwidth]{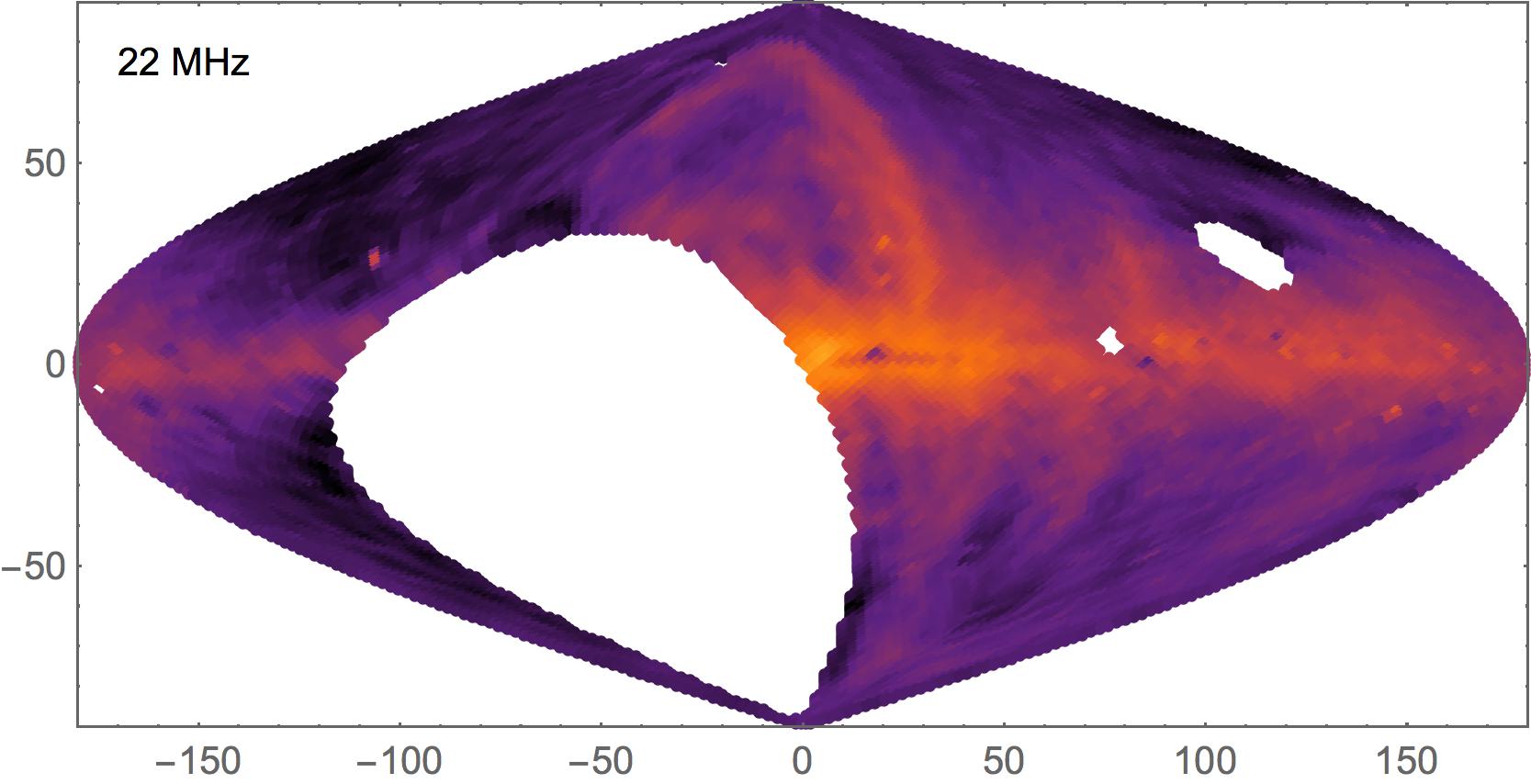} \hfill
\includegraphics[width= 0.33 \textwidth]{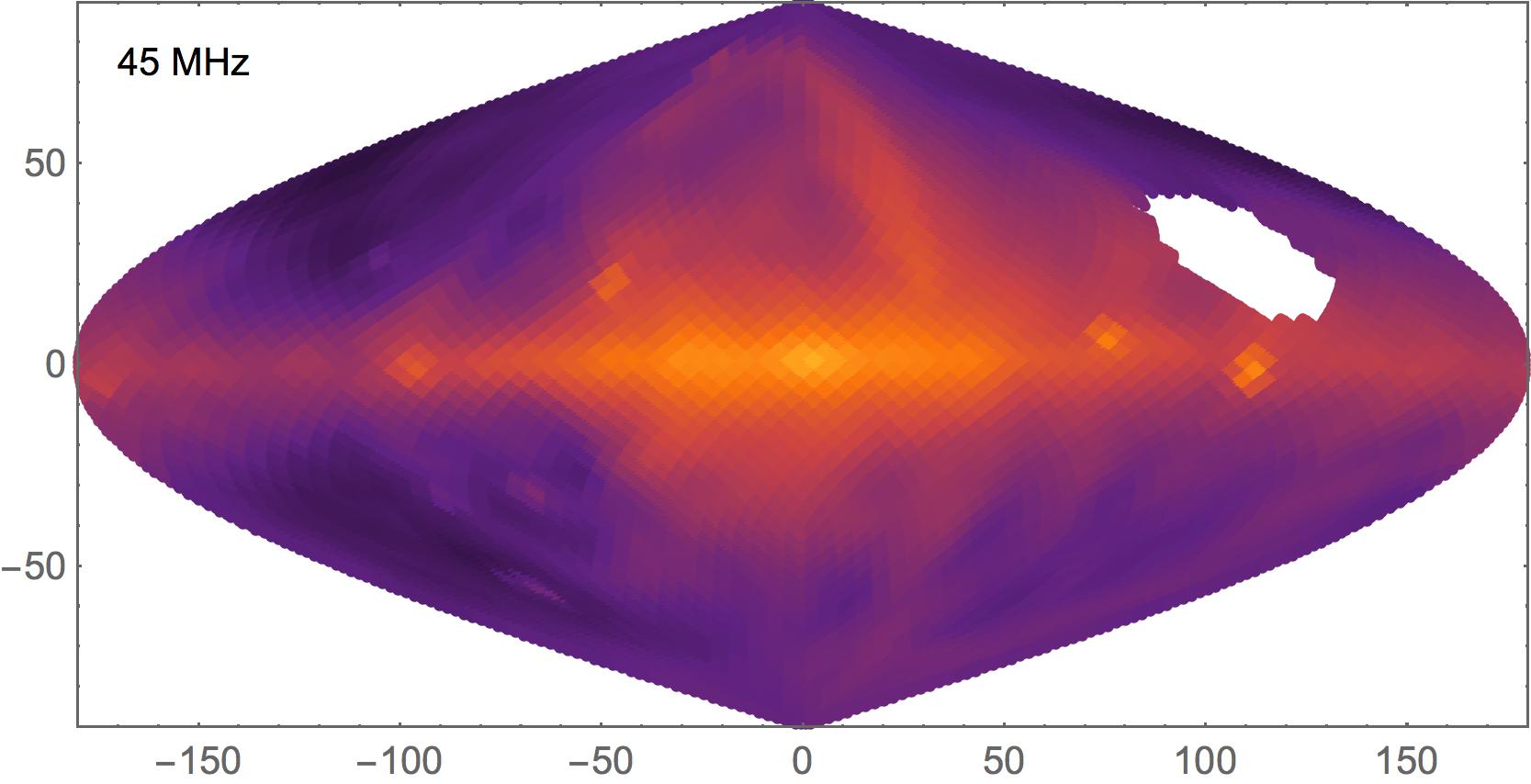}\hfill
\includegraphics[width= 0.33 \textwidth]{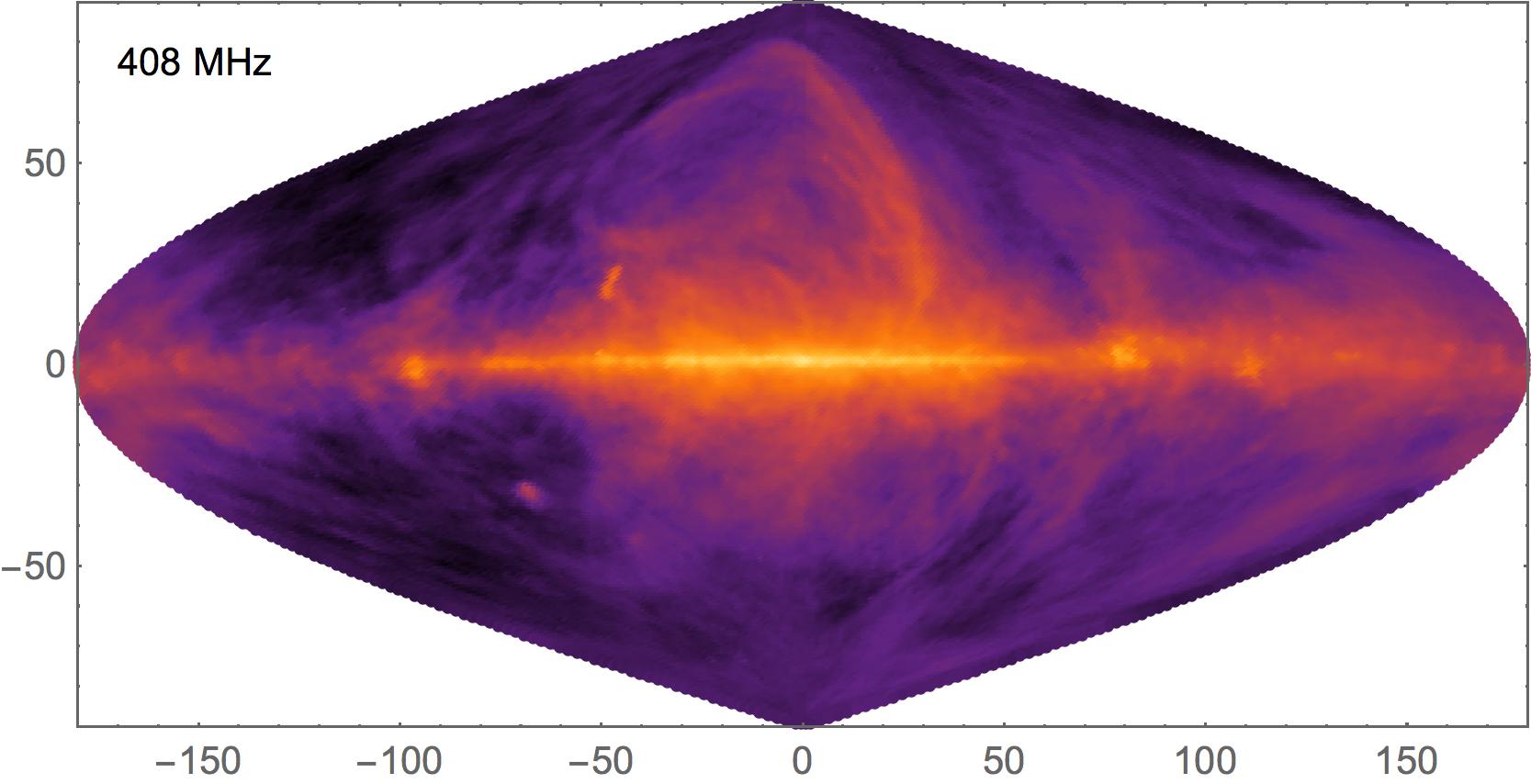} \\[0.4 cm]
\includegraphics[width= 0.33 \textwidth]{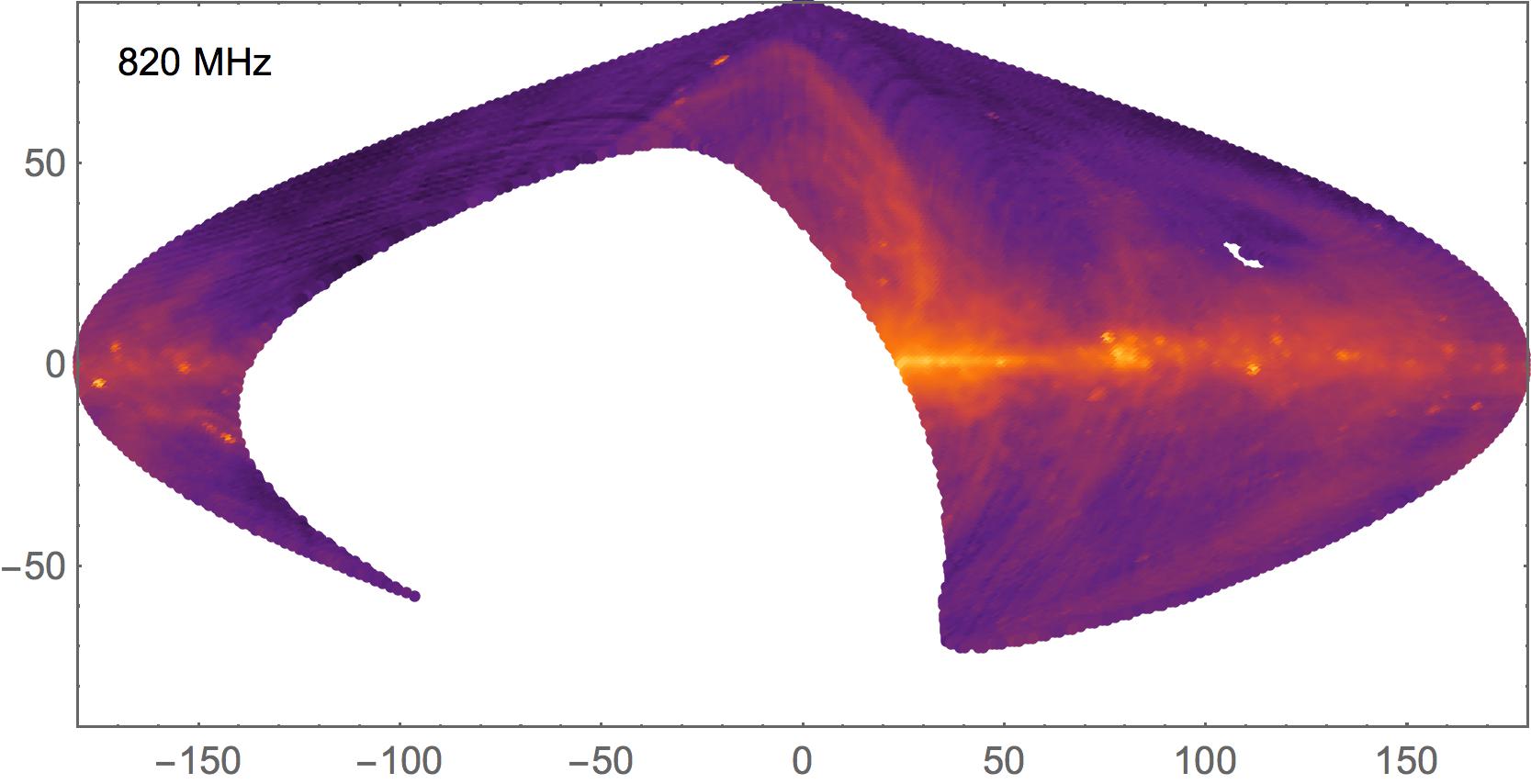}\hfill
\includegraphics[width= 0.33 \textwidth]{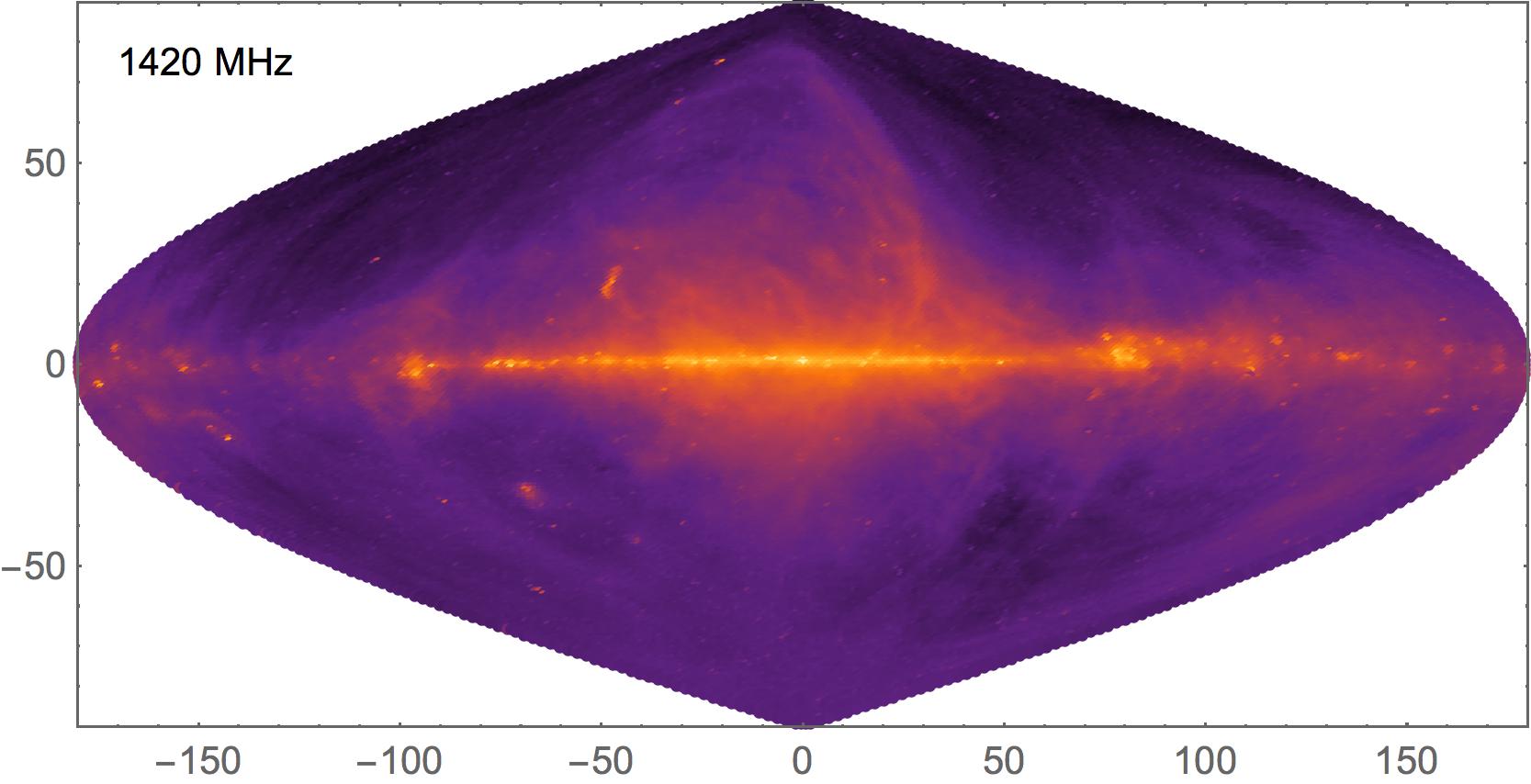}\hfill
\includegraphics[width= 0.33 \textwidth]{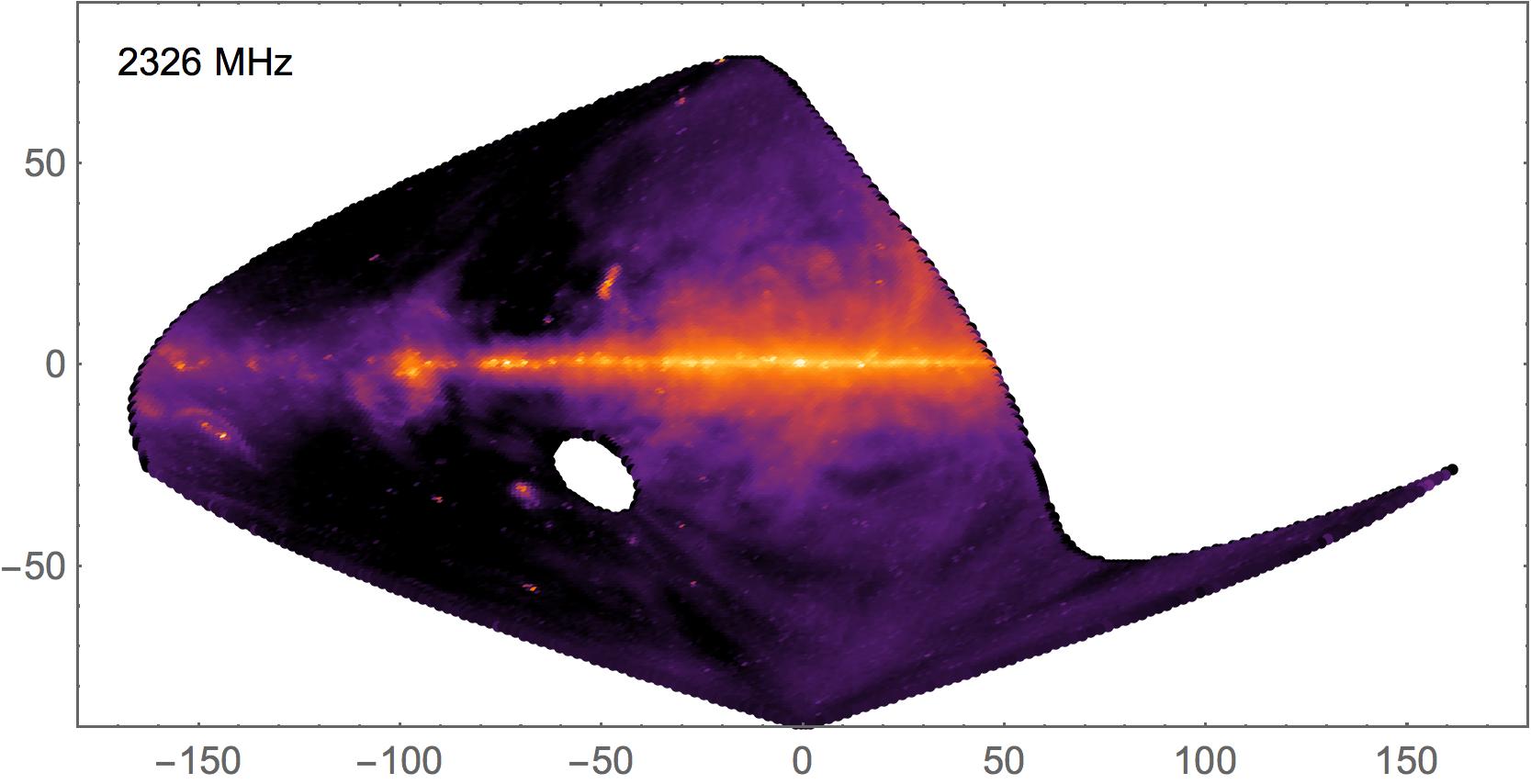} \\[0.4 cm]
\includegraphics[width= 0.33 \textwidth]{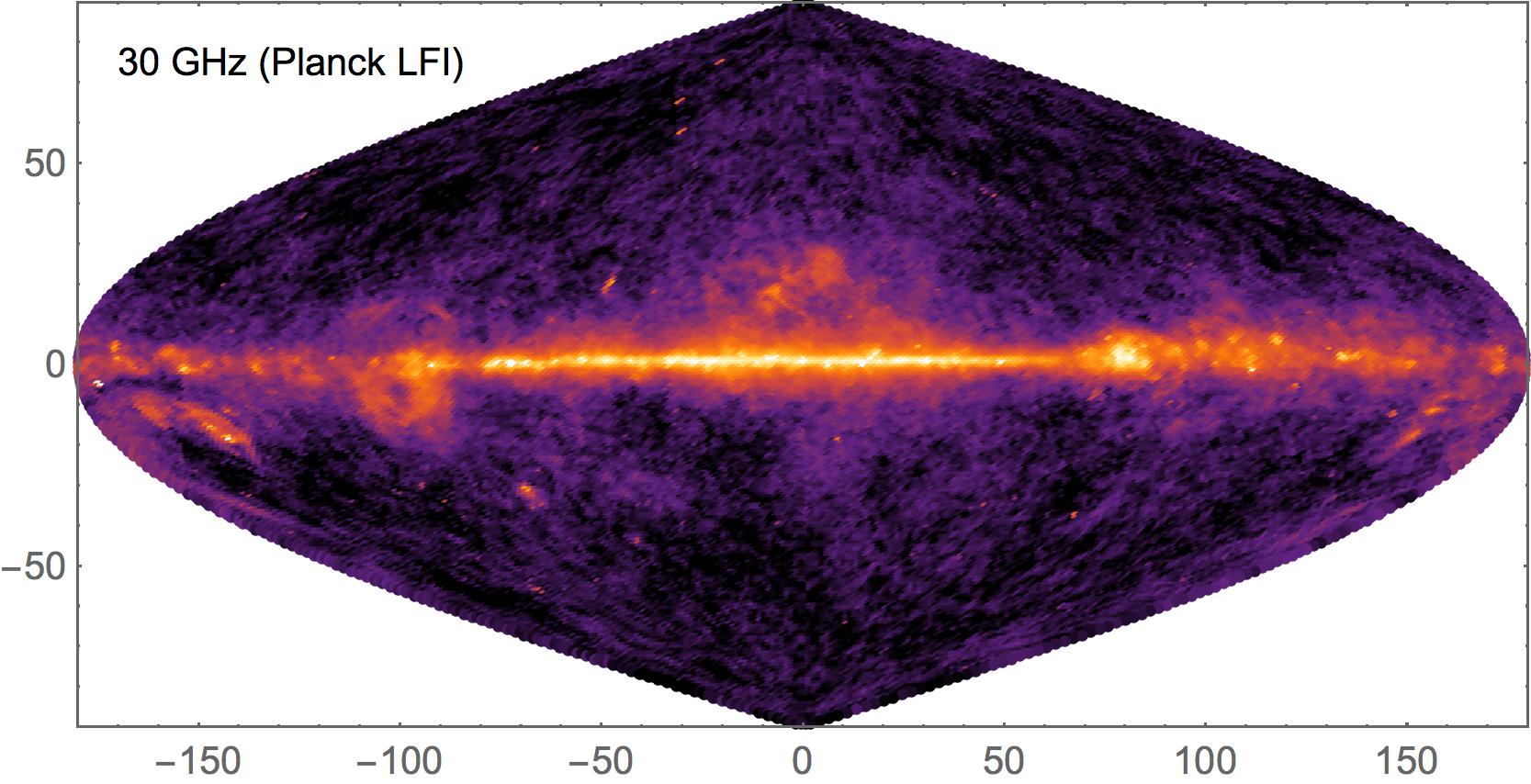}\hfill
\includegraphics[width= 0.33 \textwidth]{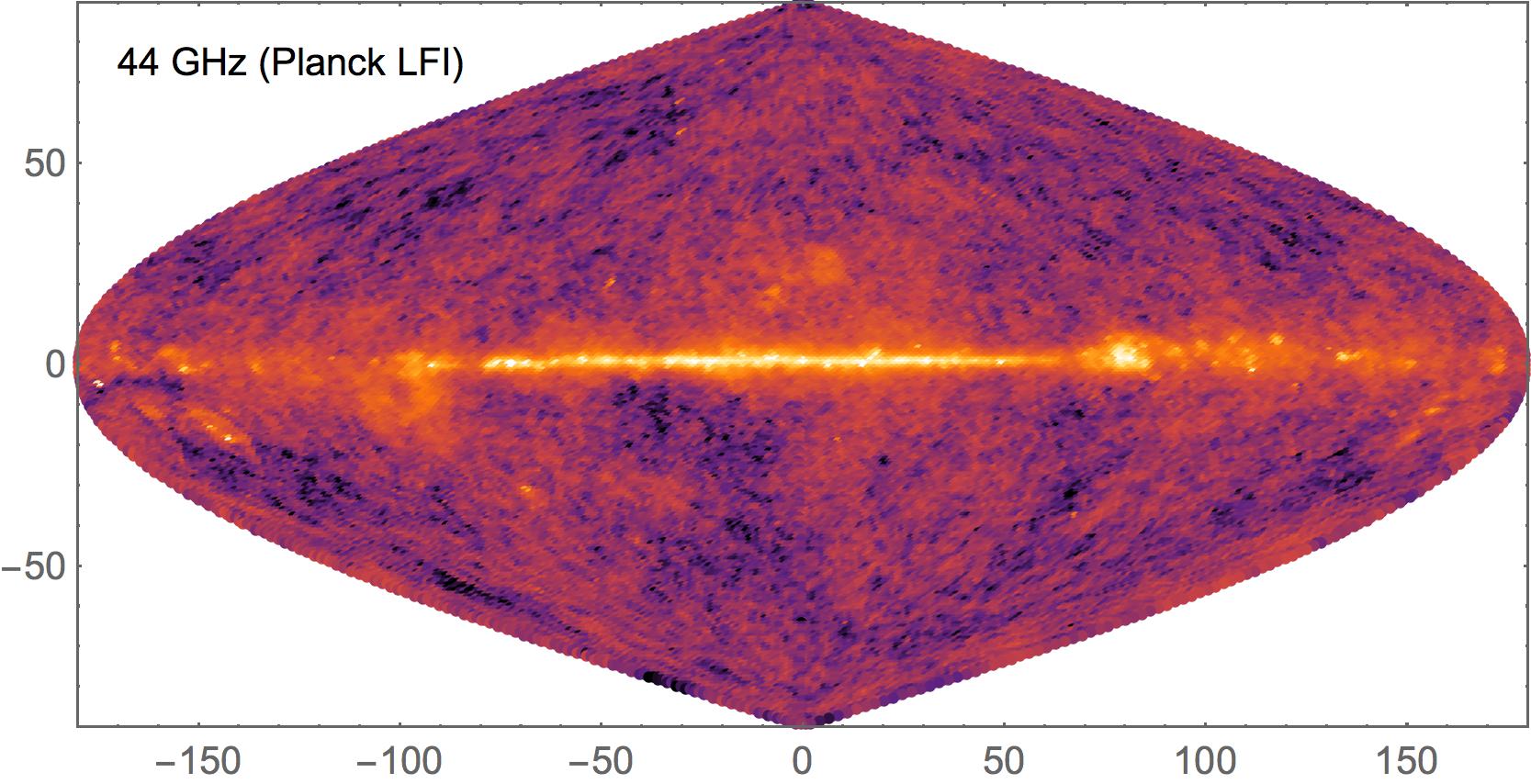}\hfill
\includegraphics[width= 0.33 \textwidth]{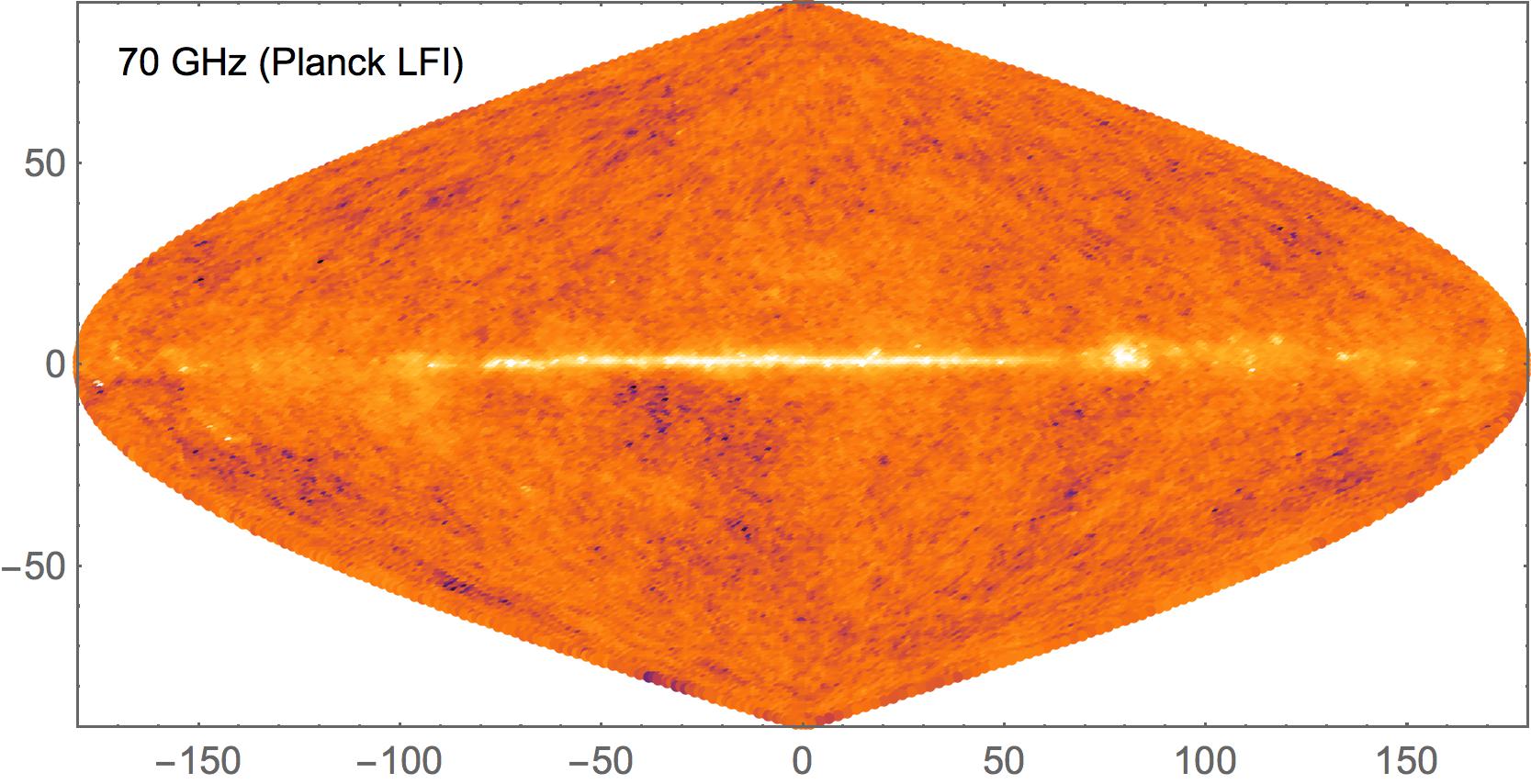}\\[0.4 cm]
\includegraphics[width= 0.33 \textwidth]{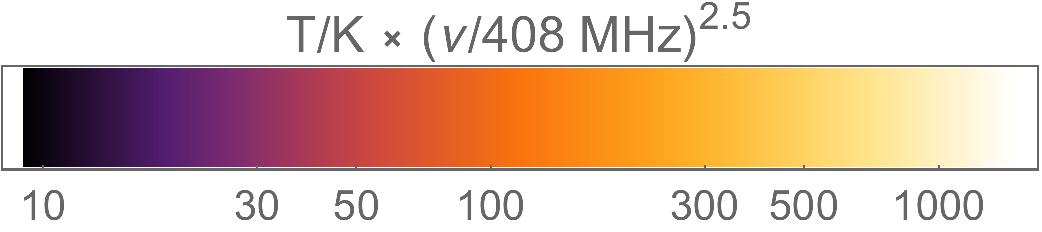}
\caption{\em \small \label{fig:maps} {\bfseries Radio maps} that we use in the analysis. The color scale is given in the legend and is common to all maps: the temperature values are as marked for $\nu$ = 408 MHz, while they have to be rescaled with $\nu^{2.5}$ for the other frequencies. From the three Planck-LFI maps the CMB monopole has been subtracted and they are offset to have a positive temperature in all pixels (see text for details). }
\end{center}
\end{figure}

On this basis we select six radio maps from 22 MHz up to 2.3 GHz. Their properties are summarized in table~\ref{tab:maps}. In ref.~\cite{Fornengo:2014mna} these maps have been converted in the HEALPix \cite{Gorski:2004by} format, with a resolution close to the original one of the surveys \footnote{The maps can be downloaded from~\cite{webmaps}.}. In addition, we take the three {\sc Planck-Lfi} maps at 30, 44 and 70 GHz.\footnote{Note that each {\sc Planck-Lfi} map is actually derived from the combined observations of several receivers operating at frequencies scattered on a (narrow) band. The central frequencies of such bands are in fact 28.4, 44.1 and 70.4 MHz~\cite{Mennella:2011ay,Ade:2015sua}. However, all {\sc Planck} publications refer to the maps conventionally as 30, 44, 70. For simplicity, we will consider the frequencies fixed and we stick to the standard nominal values.} All the maps are reproduced in fig.~\ref{fig:maps}.

\begin{table}[t]
\centering
\begin{tabular}{l|c|c}
{\em Frequency} & {\em Source and Reference} & {\em Sky coverage}  \\
\hline
22 MHz & Roger et al. ~\cite{DRAO:22} &73\% \\
45 MHz &  Guzman et al.~\cite{CHILE} & 96\% \\
408 MHz & Haslam et al.~\cite{Haslam} & 100\%  \\
820 MHz &  Berkhuijsen~\cite{DWING} &51\%\\
1420 MHz & Reich et al.~\cite{Stockert1,Stockert2,VILLA} &  100\%  \\
2326 MHz &  Jonas et al.~\cite{Jonas} &97\%\\
\hline
30 GHz & {\sc Planck-Lfi}~\cite{PLA} & 100\%  \\
44 GHz & {\sc Planck-Lfi}~\cite{PLA} & 100\%  \\
70 GHz & {\sc Planck-Lfi}~\cite{PLA} & 100\%  \\
\end{tabular}
\caption{\em \small \label{tab:maps} Details of the {\bfseries radio maps} that we use in this study.}
\end{table}

\bigskip

The use of the {\sc Planck} maps is complicated by the very presence of the CMB in them.\footnote{We comment in passing that, of course, an alternative approach, more aggressive with respect to the one we discuss below, would consist in using `foreground-only maps' (provided by the {\sc Planck} collaboration~\cite{PLA}) that have already been cleaned of the CMB. We do not pursue this strategy since we want to adopt a strategy as conservative as possible: especially for the first kind of constraints that we will present below, we want to confront the possible DM emission to the total sky brightness, rather than to individually extracted components.} For completeness, we provide here some technical comments but we anticipate that these details are not crucial for the resulting DM bounds. 
We download the {\sc Planck-lfi} temperature maps at the different relevant frequencies from the {\sc Planck} Legacy Archive~\cite{PLA}. Following the digital documentation accompanying the maps and following~\cite{Ade:2015sua}, we learn that the maps, as they have been released, have been processed to remove in particular a space-independent `zero-level' astrophysical emission (whose precise value is given in Table 12 of~\cite{Ade:2015sua} for each map) as well as, of course, the CMB monopole. We re-instate the `zero-level'~\footnote{Technically, due to the relative signs of the contributions, this actually consists, e.g.~in the case of the 30 MHz map, in {\em subtracting} $\sim 81 \ \mu K$.}, so that, to the best of our understanding, we end up having maps that contain the bulk of the astrophysical emissions as well as the CMB temperature fluctuations. Now, such maps naturally contain negative temperature pixels, especially in regions far from the Galactic plane: they correspond to spots in which the CMB has a cold fluctuation and the astrophysical intensity is not enough to compensate it. We have to offset such negative values, which are just a relic of the fact that we cannot disentangle the CMB cold spots from the whole emission. In order to be as conservative as possible, we therefore reset the minimum temperature of each map to zero.\footnote{Technically, for the case of the 30 MHz map this consists in adding $\sim 305 \ \mu K$ overall.} This is tantamount to adding to the map an {\em effective uncertainty} that corresponds to the deepest negative temperature fluctuation. We have checked that adding a larger uncertainty (e.g. twice the size of the deepest negative fluctuation) impacts the bounds by a negligible amount. This is not surprising: as we will see later, the dominant DM bounds come from regions close to the GC, in which the added uncertainty is negligible compared to the astrophysical emission.


\section{Galactic radio emission from Dark Matter} 
\label{sec:formalism}

As mentioned above, DM produces radio and microwave signals as secondary radiation, in the sense that these signals are emitted (via the synchrotron process) by the electrons and positrons produced in the annihilation or decay process, as they gyrate in the environmental magnetic field. Here we review the basics of this process. For all details we refer to~\cite{Buch:2015iya}, that we use extensively for all the computations. 

\medskip

The DM synchrotron intensity $\mathfrak{I}$ at a frequency $\nu$ from a given line of sight (individuated by the galactic coordinates $b,\ell$) is given by 
 
 \begin{equation}
\mathfrak{I}(\nu,b,\ell) = \frac{1}{4\pi} \int_{\rm l.o.s.}ds \  \int_{m_e}^{M_{\rm DM}(/2)} dE\ 2\, \mathcal{P}_{\rm syn}(\nu,E)\,  f(E,r,z) .
\label{eq:sync}
\end{equation}

\noindent The first portion just consist of the spatial integral along the line of sight (parameterized by the coordinate $s$). The second integral is the synchrotron emissivity of a cell located at a position $(r,z)$ along the line of sight. It is expressed as the convolution of two quantities: the number density $f$ of electrons or positrons in $(r,z)$ and $\mathcal{P}_{\rm syn}$, the synchrotron power emitted at the frequency $\nu$ by an electron with energy $E$. All DM-produced electrons (and positrons) possessing an energy between $m_e$ (the electron mass) and the maximum one $M_{\rm DM}$ or $M_{\rm DM}/2$ (the DM particle mass for the annihilation case or half of it for the decay case, respectively) contribute to the intensity. The factor 2 in front of the power accounts for the presence of an equal population of electrons and positrons. The number density $f$ of emitting $e^\pm$ results from the computation of the DM injection, via the annihilation or decay processes, plus the subsequent propagation in the galactic environment. It is determined therefore on the basis of the information about the DM $e^\pm$ spectra ($dN^{f}_{e^\pm}/dE$) in a given annihilation or decay channel $f$, the annihilation cross section $\langle \sigma v \rangle_f$ or the decay rate $\Gamma_f$ in the same channel, the spatial distribution of DM in the Galaxy and the parameters controlling the $e^\pm$ propagation, including the intensity and distribution of the magnetic field. The power $\mathcal{P}_{\rm syn}$ also depends on the magnetic field configuration, of course. 
 
\medskip

What is sketched above is the {\em ab initio} computation of the synchrotron intensity from DM. In the practical computations, however, it is very convenient to recast the formul\ae\ in terms of a convolution of the $e^\pm$ injected by DM and a set of generalized synchrotron halo functions $I_{\rm syn}$ which encapsulate all the astrophysical ingredients~\cite{Buch:2015iya}. Namely, one has

\begin{equation}
 \mathfrak{I}(\nu,\ell,b)=\frac{r_\odot}{4\pi} \left\{
 \begin{array}{l}
 \displaystyle\!\!
 \frac{1}{2}\left(\frac{\rho_{\odot}}{M_{\rm DM}}\right)^2  \int_{m_e}^{M_{\rm DM}} dE_s  \sum_f \langle \sigma v \rangle_f \frac{dN^f_{e^{\pm}}}{dE}(E_s)\ I_{\rm syn}(E_s,\nu,\ell,b) \quad {\rm (annihilation)}\\
\displaystyle\!\!
 \left(\frac{\rho_{\odot}}{M_{\rm DM}}\right) \int_{m_e}^{M_{\rm DM}/2} dE_s  \sum_f \Gamma_f \frac{dN^f_{e^{\pm}}}{dE}(E_s)\ I_{\rm syn}(E_s,\nu,\ell,b) \quad {\rm (decay)}
 \end{array}
 \right.
\label{eq:synchintensity}
\end{equation}
where the factors of $\rho_{\odot}$ (the position of the Sun with respect to the Galactic Center) and $\rho_\odot$ (the local DM density) are just introduced for normalization purposes. The functions $I_{\rm syn}$ are provided by~\cite{Buch:2015iya} while the $dN^{f}_{e^\pm}/dE$ spectra are provided by~\cite{Cirelli:2010xx} (and~\cite{Ciafaloni:2010ti}), both in the context of the {\sc Pppc4dmid} package~\cite{website}.

Finally, the synchrotron intensity $\mathfrak{I}$ is traded for the brightness temperature $T$, espressed in Kelvin, which is just the temperature that a black body would have to possess in order to emit the same intensity at a given frequency. One has
\begin{equation}
 T(\nu)=\frac{c^2 \, \mathfrak{I}(\nu)}{2\, \nu^2 \, k_B} ,
\label{eq:syncT}
\end{equation}
with $k_B$ the Boltzmann constant.

\medskip

With the above technology we can compute DM synchrotron temperature spectra and maps, to be directly compared with surveys, for each choice of the particle physics and astrophysics variables.

Concerning the annihilation/decay channels, we consider eight representative ones: $e^+e^-$, $\mu^+\mu^-$, $\tau^+\tau^-$, $b \bar b$, $t \bar t$, $W^+W^-$, $hh$ and $\gamma\gamma$. We have therefore three leptonic channels, two (heavy) quark channels, a weak gauge boson channel (the $ZZ$ channel would be essentially indistinguishable), the higgs boson channel and a direct $\gamma\gamma$ channel. While the generic expectation would be that the leptonic channels produce abundant high-energy $e^\pm$ (hence the largest synchrotron signal) and that, e.g., the $\gamma\gamma$ channel do not produce any, we will see below that the electroweak corrections (and other effects) partly modify these features. 

Concerning the DM galactic distribution, we consider the standard NFW~\cite{Navarro:1995iw}, Einasto~\cite{Graham:2005xx, Navarro:2008kc} and Burkert~\cite{Burkert} profiles. While the former two are peaked at the GC, the latter one exhibits a core. The precise parameters defining the profiles are reported in~\cite{Buch:2015iya}.

Concerning the $e^\pm$ propagation parameters in the Galaxy, we use the conventional {\sc Min}, {\sc Med}, {\sc Max} set~\cite{DonatoPRD69,FornengoDec2007}. The precise values and their meaning are again reported, e.g., in~\cite{Buch:2015iya}. 
More refined diffusion/propagation schemes might be employed (e.g.~to include effects such as diffusive reacceleration or a spatially varying diffusion coefficient), but they are beyond the scope of our analysis.

Concerning the magnetic field in the Galaxy, we adopt the {\sc Mf1}, {\sc Mf2}, {\sc Mf3} configurations discussed in~\cite{Buch:2015iya}. They all carry a double-exponential dependance on $r$ and $z$. {\sc Mf1} is an often-used configuration, {\sc Mf2} extends less in $z$ while {\sc Mf3} extends further. In terms of intensity at GC, they are all contained between 10 and 15 $\mu$G .
These parametrizations are intended to capture the basic features of more complex and realistic magnetic field models, like those in~\cite{Jansson:2012rt,Jansson:2012pc,Sun2008,Sun:2010sm,  Pshirkov:2011um}. Employing these detailed magnetic fields would have an impact on the synchrotron signal computed here. Still, since we focus on large  scale emissions, we expect that these changes would  largely be reabsorbed by the variation of  the magnetic field spanned by our {\sc Mf1, Mf2, Mf3} models.

\begin{figure}[t]
\begin{center}
\includegraphics[width= 0.32 \textwidth]{figures/408.jpg} \hfill
\includegraphics[width= 0.32 \textwidth]{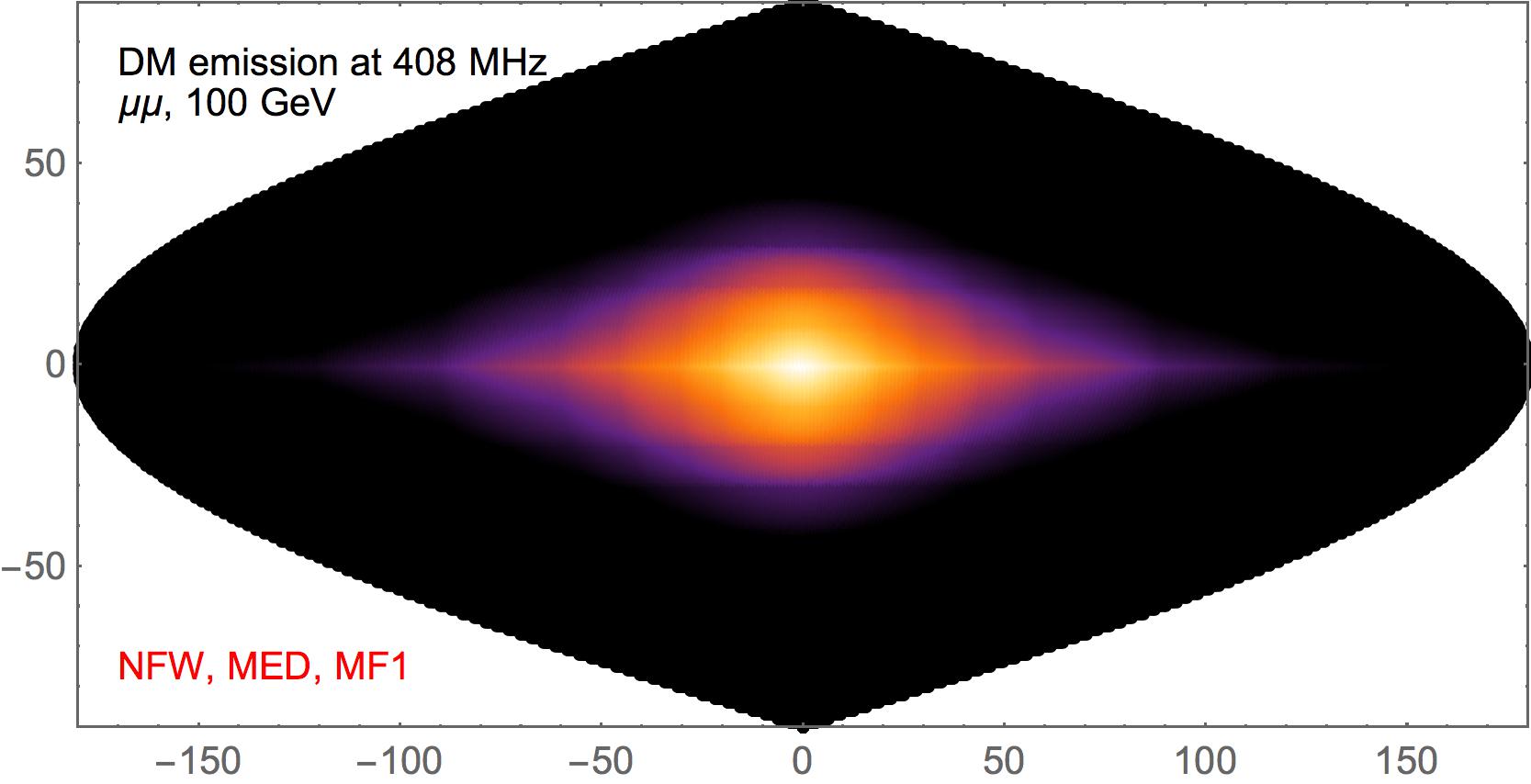} \hfill
\includegraphics[width= 0.32 \textwidth]{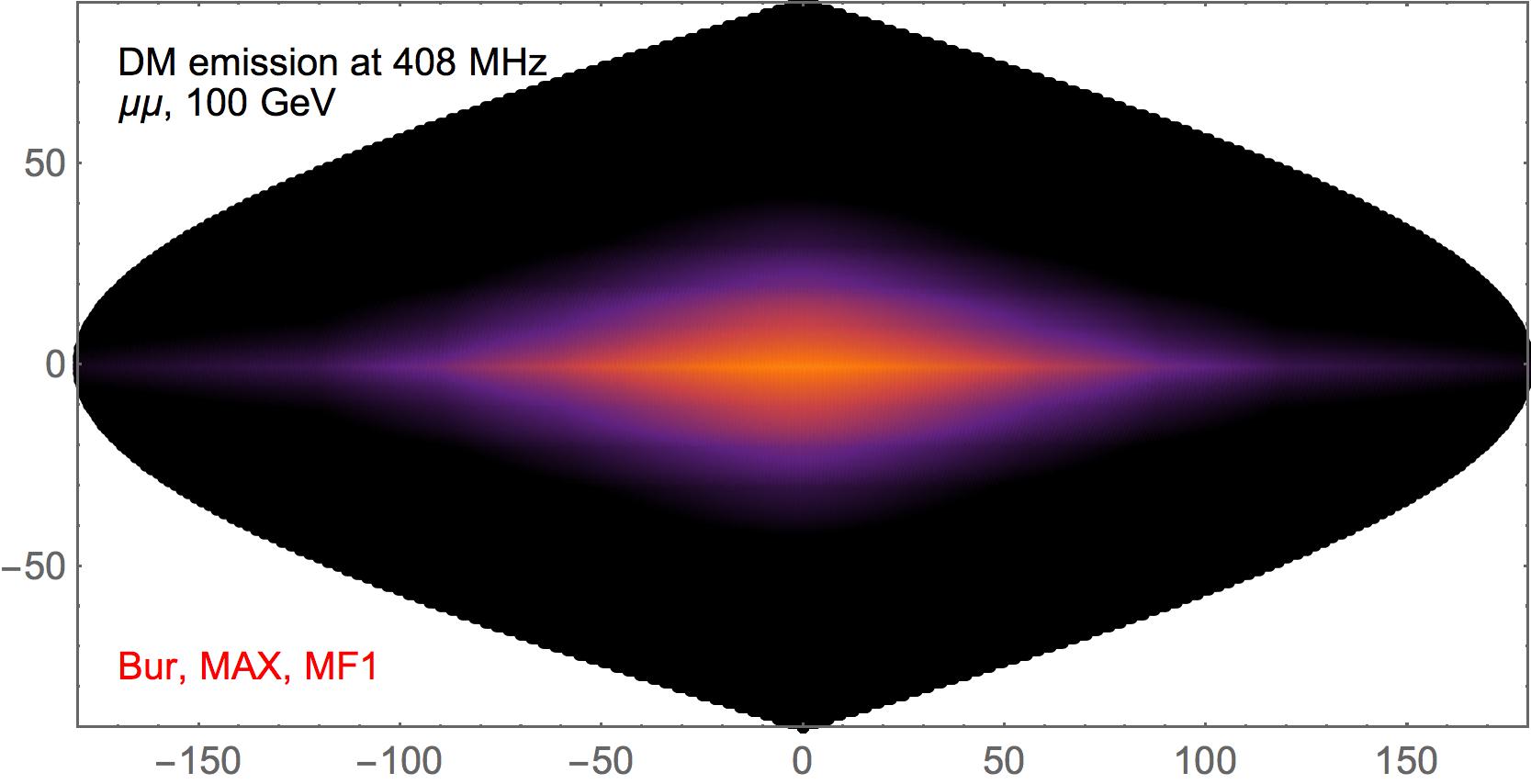} \\[0.4cm]
\includegraphics[width= 0.32 \textwidth]{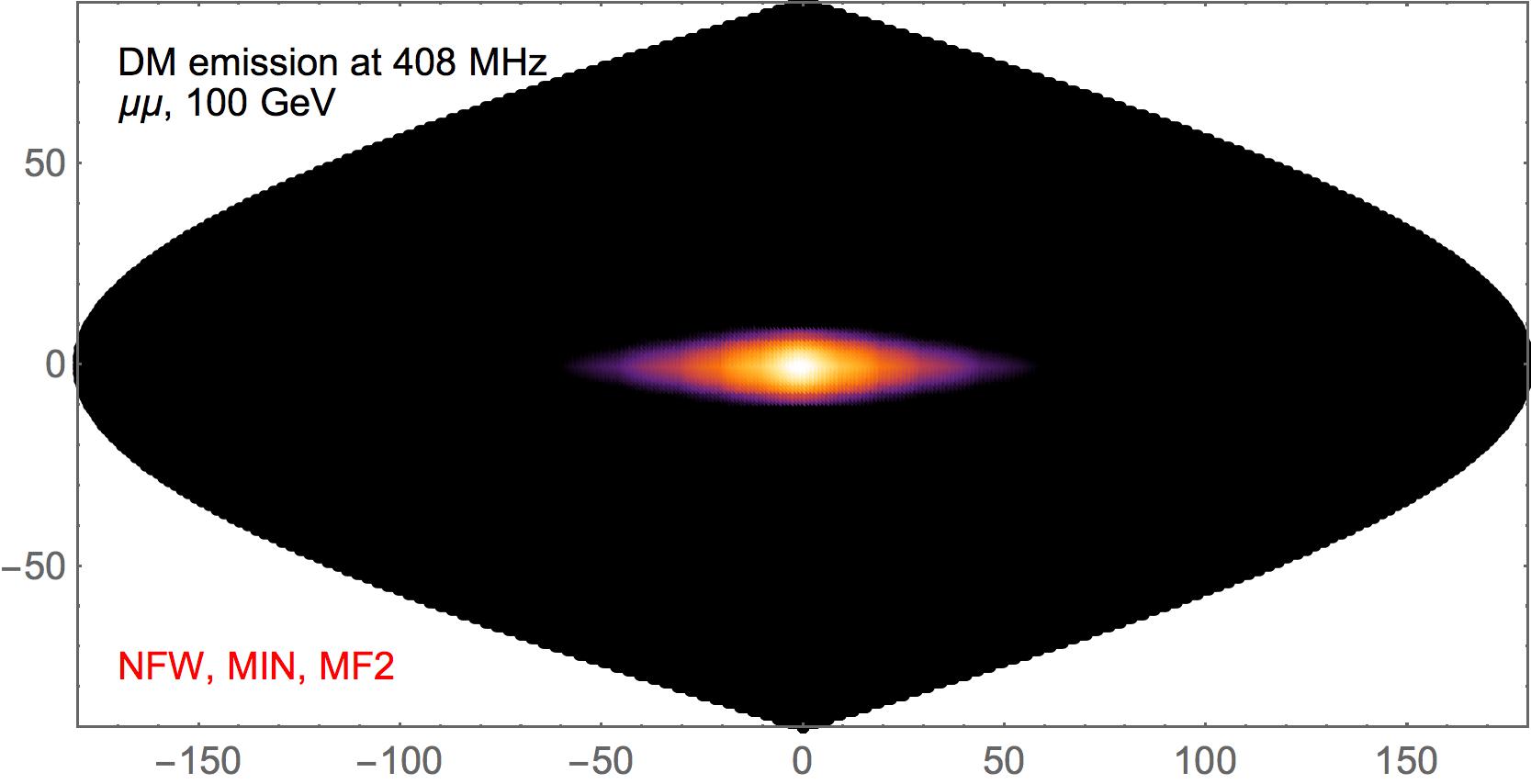} \hfill
\includegraphics[width= 0.32 \textwidth]{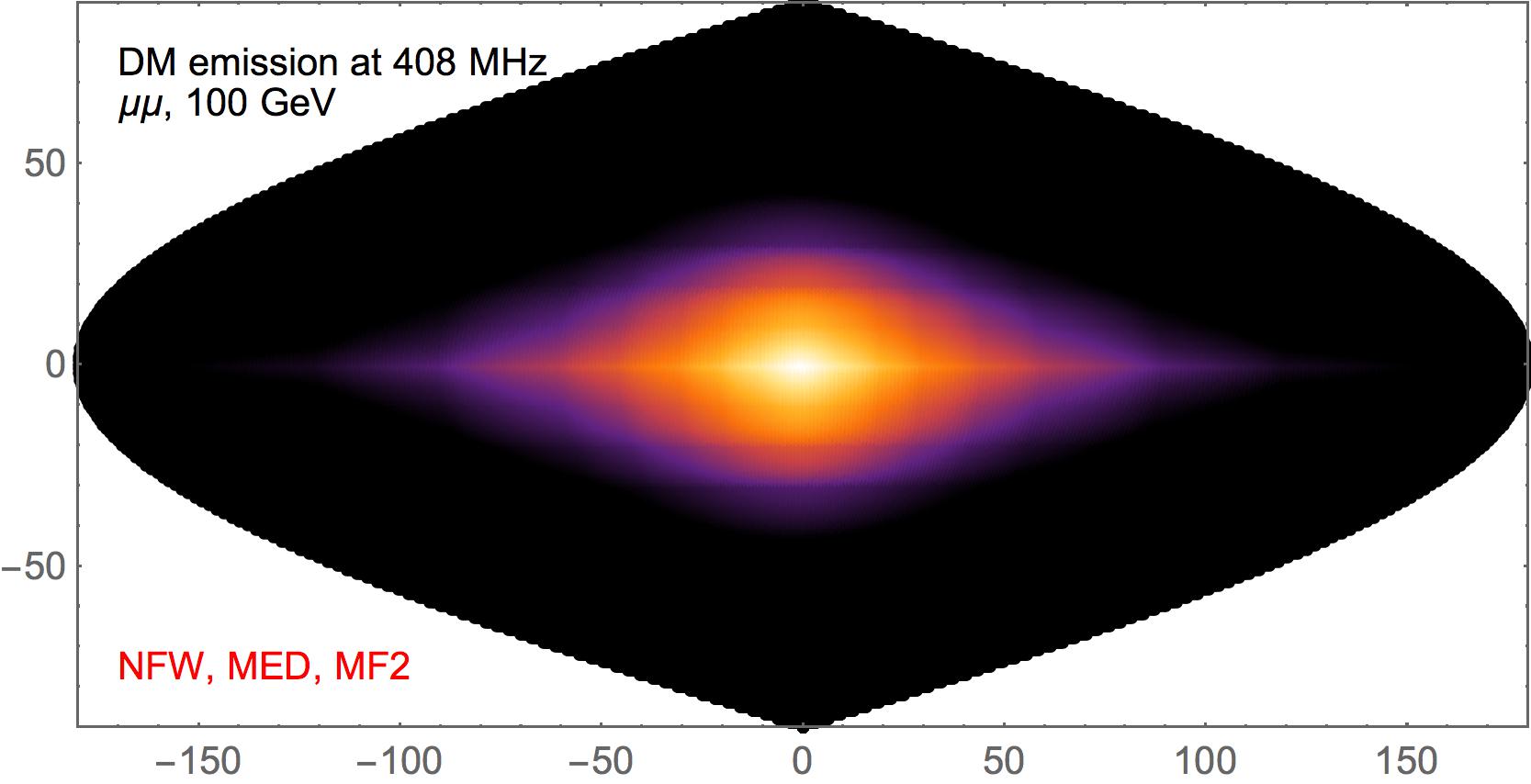} \hfill
\includegraphics[width= 0.32 \textwidth]{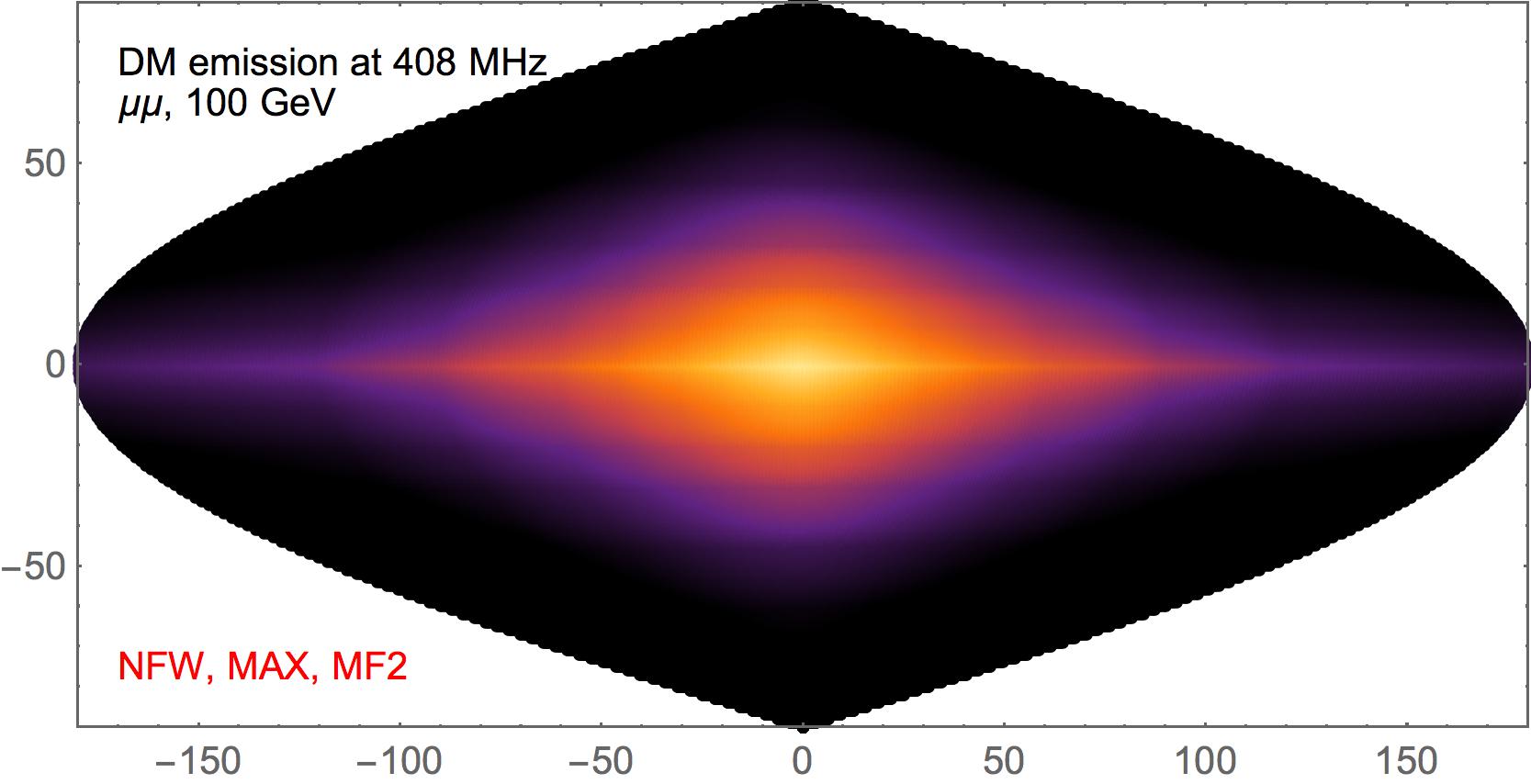} \\[0.4cm]
\includegraphics[width= 0.32 \textwidth]{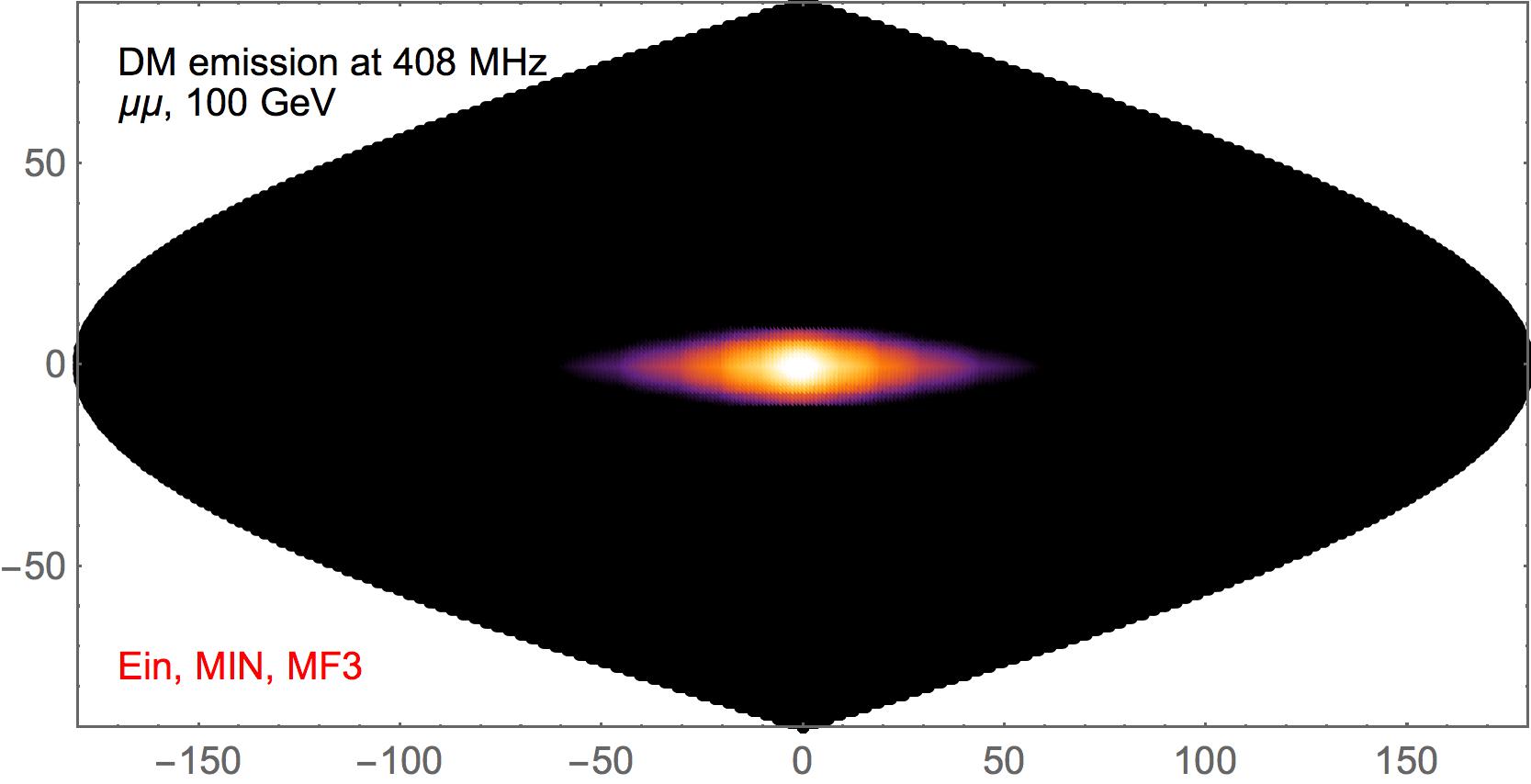} \hfill
\includegraphics[width= 0.32 \textwidth]{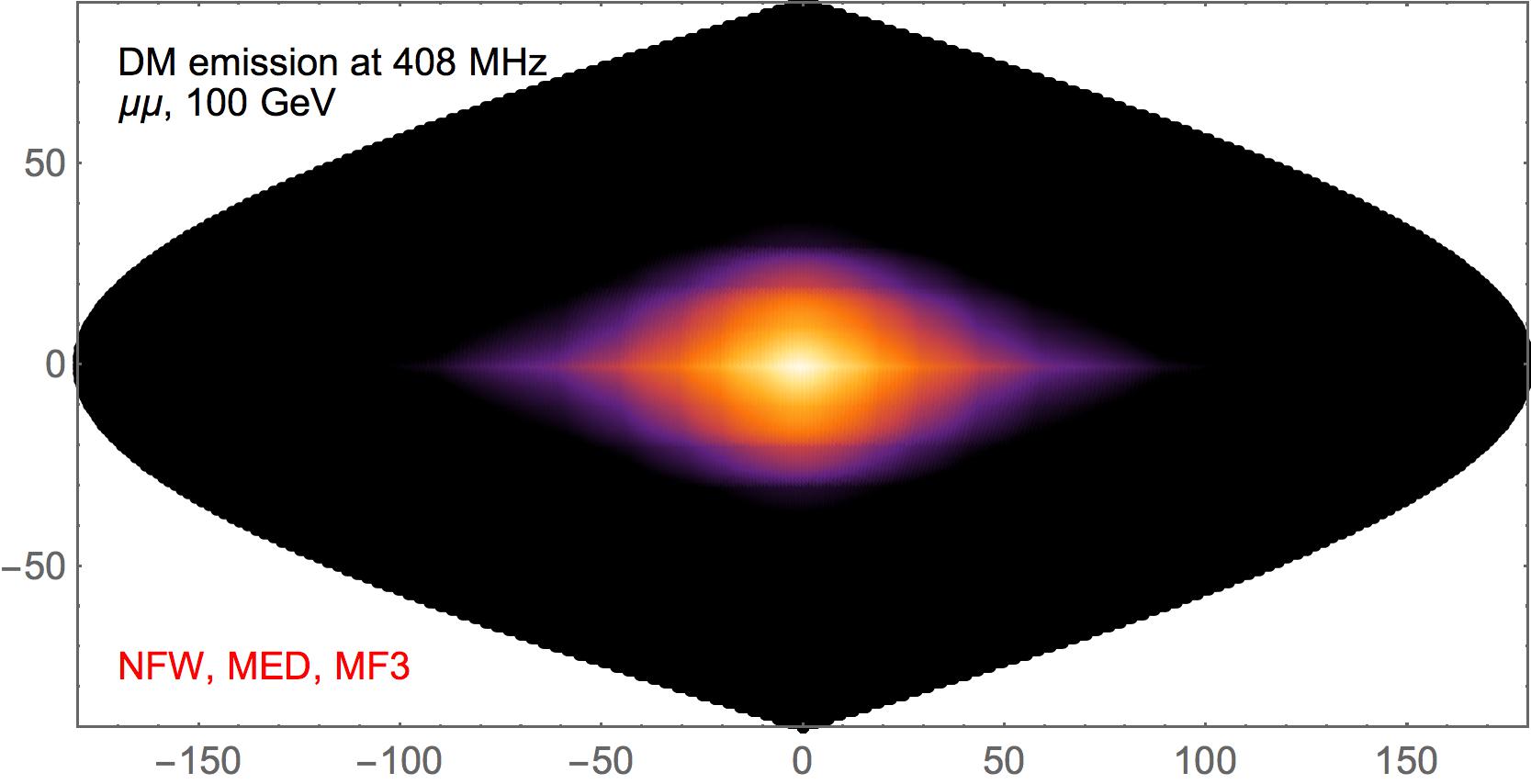} \hfill
\includegraphics[width= 0.32 \textwidth]{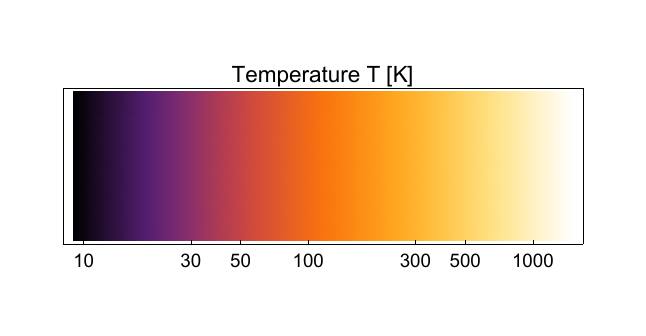} \\[0.4cm]
\caption{\em \small \label{fig:DMmaps} {\bfseries Annihilating DM radio emission maps} for different cases, together with the observed sky at the same frequency (top left). Each column from left to right: MIN, MED, MAX. Each line from top to bottom: MF1, MF2, MF3. The DM profile is always NFW, except for top-right (Burkert) and bottom-left (Einasto). The DM annihilation cross section is fixed at the large value of $\langle \sigma v \rangle = 3\, 10^{-23} \, {\rm cm}^3/{\rm s}$ in order to have a brightness comparable to that of the observed sky, for better illustration.}
\end{center}
\end{figure}

\medskip

In fig.~\ref{fig:DMmaps} we present some DM synchrotron temperature maps at 408 MHz, varying some of the parameters discussed above. For definiteness, we choose the $\mu^+\mu^-$ annihilation channel, a DM mass of 100 GeV and fix the annihilation cross section at the arbitrary value of $3 \ 10^{-23}$ cm$^3$/s, in order to have a brightness comparable to the one of the observed sky at the same frequency. By eye, we see that changing the propagation parameters {\sc Min, Med, Max} (from left to right in the middle row of panels in the figure) implies a change in the morphology of the signal: for {\sc Min} the emission is more intense at the GC and does not extend at all at large latitudes, for {\sc Max} a good portion of the sky is illuminated by DM. This is easily understandable in terms of the fact that {\sc Max} features a thick diffusive halo, that contains the radiating $e^\pm$ up to rather large latitudes. Changing the magnetic field configuration {\sc Mf1, Mf2, Mf3} (central column of panels) has a very limited impact. 
Somewhat counterintuitively, {\sc Mf3} features a slightly less extended and less intense emission. This is due to the complicated interplay between energy losses and emission during the $e^\pm$ propagation. Indeed, a stronger magnetic field (such as {\sc Mf3} is, with respect to the other ones) implies larger energy losses for the electrons and positrons, so that their steady state population is comparatively depressed and emits less. 
The differences are however very small. Finally, changing the DM distribution among NFW, Einasto and Burkert has a more relevant impact: the cored profile (top right panel in fig.~\ref{fig:DMmaps}) features an emission which is significantly dimmer and more diffuse than the peaked profiles, since it largely traces the distribution of the DM source. We will recognize the consequences of these features in the bounds derived in sec.~\ref{sec:constraints}.

\medskip

\begin{figure}[p]
\thisfloatpagestyle{empty}
\begin{center}
\includegraphics[width= 0.35 \textwidth]{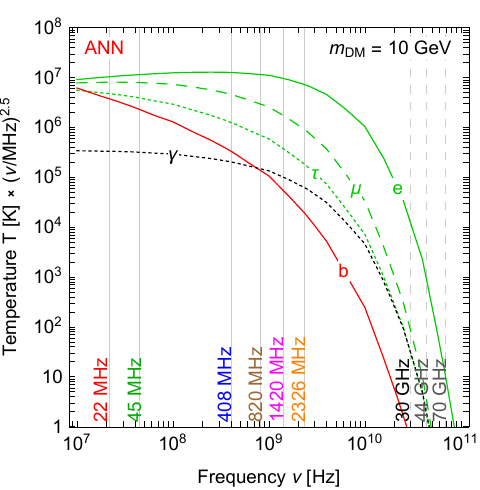} \quad
\includegraphics[width= 0.35 \textwidth]{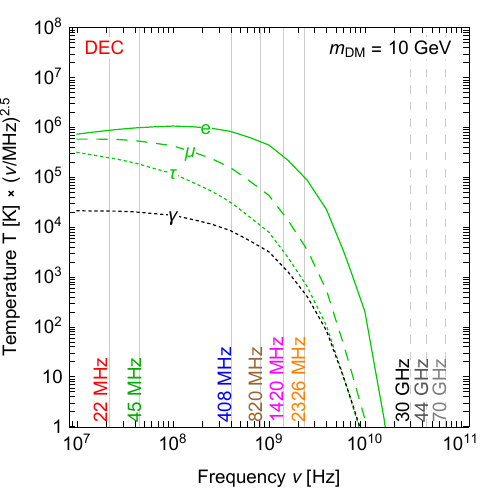}\\
\includegraphics[width= 0.35 \textwidth]{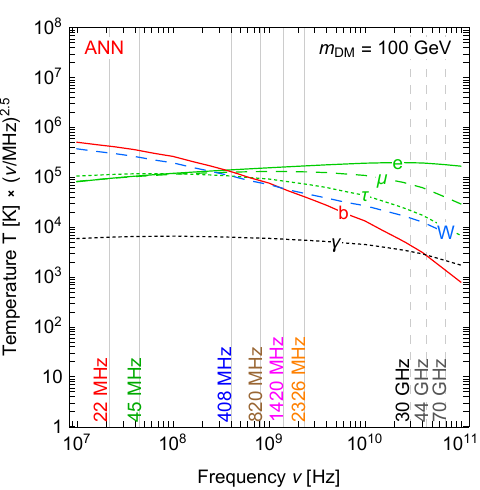} \quad
\includegraphics[width= 0.35 \textwidth]{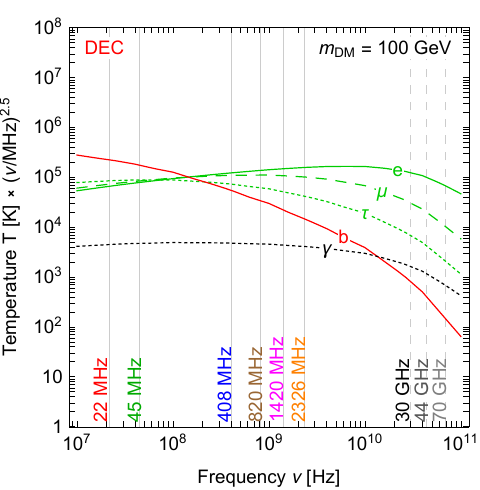}\\
\includegraphics[width= 0.35 \textwidth]{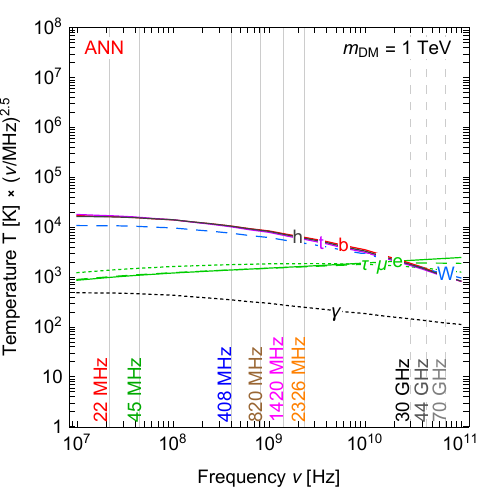}\quad
\includegraphics[width= 0.35 \textwidth]{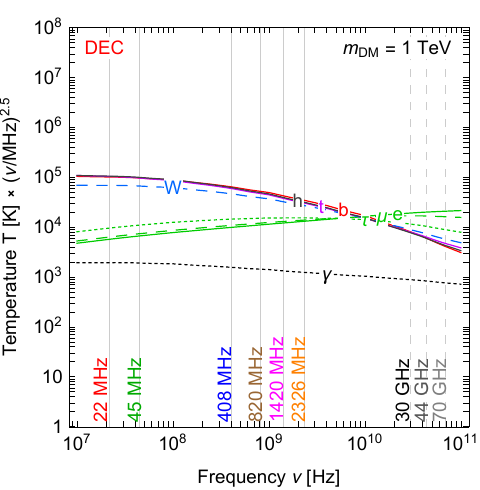}\\
\includegraphics[width= 0.35 \textwidth]{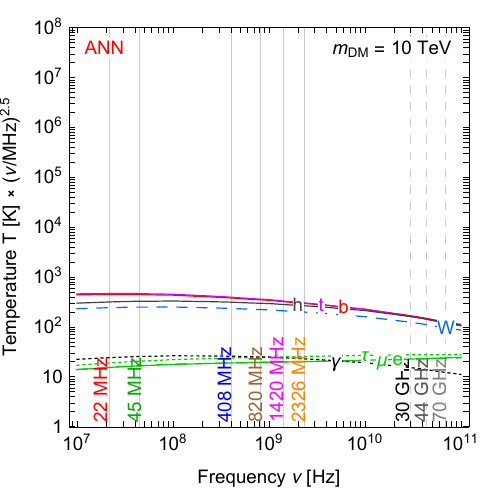}\quad
\includegraphics[width= 0.35 \textwidth]{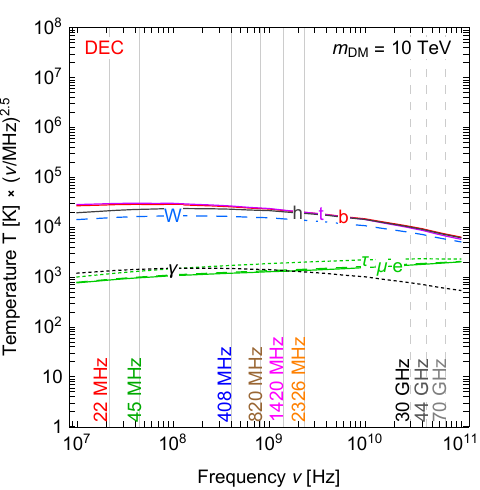}
\caption{\em \small \label{fig:DMTspectra} {\bfseries  Synchrotron radiation emission from Dark Matter} annihilations (left column) or DM decay (right column), for different channels. The vertical lines mark the frequencies of the maps we use. In all cases we assume an {\em NFW} profile, {\em MED} propagation parameters , {\em MF1} magnetic field and $\langle \sigma v \rangle = 3 \ 10^{-26}\, {\rm cm}^3/{\rm s}$ (for the annihilation case) or $\tau = 10^{28}\,{\rm s}$ (for the decay case). The line of sight is chosen to be $(\ell,b)=(20^\circ, 20^\circ)$.}
\end{center}
\end{figure}

\begin{figure}[t]
\begin{center}
\includegraphics[width= 0.47 \textwidth]{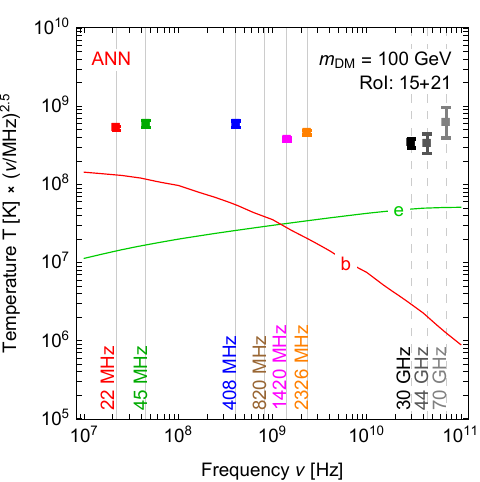} \hfill
\includegraphics[width= 0.47 \textwidth]{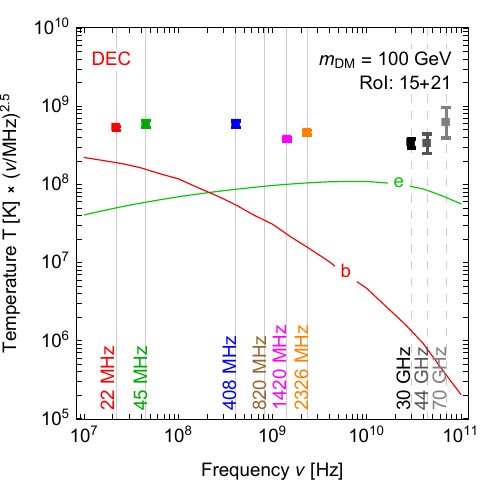} \\
\caption{\em \small \label{fig:DMTspectradata} {\bfseries Synchrotron radiation emission from Dark Matter}, this time compared with observations (colored data points). Notice that there is no measurement at 820 MHz since that map does not cover the RoI considered here. The error of the measurement is often invisible within the size of the data point. Note that, for this figure, we fix $\langle \sigma v \rangle = 10^{-25}\, {\rm cm}^3/{\rm s}$ and $\tau = 10^{26}\,{\rm s}$.}
\end{center}
\end{figure}

Let us now move to discuss the frequency dependence of the DM emission. In fig.~\ref{fig:DMTspectra} and~\ref{fig:DMTspectradata} we plot the DM synchrotron temperature as a function of $\nu$, for a chosen line of sight ($(\ell, b) = (20^\circ,20^\circ)$, for definiteness). More precisely, for reasons that will be clear in a moment, we plot the temperature multiplied by a factor $\nu^{2.5}$. Several remarks are in order. 
\begin{itemize}
\item[$\diamond$] First, let us point out that, perhaps surprisingly, the $\gamma\gamma$ channel can contribute a non-negligible synchrotron radiation. Indeed, $e^\pm$ can be produced in this channel by the splitting of the final state $\gamma$-rays. For heavy DM, the emission is as large as those of the leptonic channels.
\item[$\diamond$] Secondly, we observe that the spectrum has a cutoff in frequency, i.e. light DM cannot produce high frequency radiation (this is apparent in the first row of panels which refer to 10 GeV DM; the cutoff falls beyond our frequency range for large DM masses). Hence we anticipate that the constraints on light DM will be dominated by low frequency maps.
\item[$\diamond$] Thirdly, we notice that, for relatively low DM masses (e.g. 10 and 100 GeV), the leptonic channels feature a significantly harder spectrum than the hadronic ones, with the $e^+e^-$ one being particularly steep; the $W^+W^-$ case is somewhat intermediate. As one moves to heavy DM, however, the feature is washed out. This is because for heavier masses the more complex showering process, which includes electromagnetic and weak radiative corrections, produces $e^\pm$ spectra that are more similar across the different channels. For an explicit illustration of this point, one can see fig.~3 of \cite{Cirelli:2010xx} (top row): for 1 TeV DM, all the $e^\pm$ spectra from the different channels present a pronounced shoulder at $x \equiv E/M_{\rm DM} \sim 10^{-3}$, as a consequence in particular of radiative corrections. Hence the resulting synchrotron spectra will be self-similar. 
\item[$\diamond$] Fourthly, let us comment on the impact of the frequency dependence of the different channels. In order to do that, we compare the predicted DM emission with the data in fig.~\ref{fig:DMTspectradata}. We restrict for clarity to two channels only, and we consider for definiteness the emission in a Region of Interest corresponding to $2^\circ < b < 5^\circ, -5^\circ < \ell < +5^\circ$ (denoted as RoI 15 in sec.~\ref{sec:RoIs}), plus the symmetric region with respect to the GC. The plot shows that the measured $T \times \nu^{2.5}$ (colored data points) is rather flat as a function of the frequency $\nu$ (which, by the way, explains why it is useful to consider that specific combination of quantities). Since the DM emission, for relatively small DM masses, is not flat, the various frequencies will constraint differently the different channels. Namely, hadronic channels will be mostly constrained by low frequency maps, while leptonic channels will be dominantly constrained by high frequency maps. For higher masses, the spectral shape flattens thus all maps are in principle expected to be equally relevant. In other words, the {\sc Planck-Lfi} maps may have an important constraining power for leptonic channels, for not too large DM masses.
\item[$\diamond$] Finally, let us notice that the same qualitative features discussed in the previous points are also present in the decaying DM case (right column of plots in fig.~\ref{fig:DMTspectra} and right panel in fig.~\ref{fig:DMTspectradata}). A difference with the annihilation case, however, is in the slower suppression of the intensity with increasing $M_{\rm DM}$. This is expected: the signal scales with the inverse of the mass for the decay case, as opposed to $1/M_{\rm DM}^2$ for the annihilation case (see e.g.~eq.~(\ref{eq:synchintensity})). Thus, one can anticipate bounds which are flatter in $M_{\rm DM}$ for decays compared to annihilations.
\end{itemize}

\noindent Before moving on, we warn the reader that we are not including the effect of free-free absorption.~\footnote{We thank Oscar Macias for bringing this point to our attention.} Including it would have the effect of reducing the synchrotron signal at low frequencies (below $\sim 10^8$ Hz) and only for low latitudes ($b \lesssim 10^\circ$). We leave this possible improvement for future work.
We now move to discuss the astrophysical contribution to the galactic radio emission.


\section{Modelling the galactic radio emission from astrophysics} 
\label{sec:background}

The emission observed in the radio and microwave bands is the results of  different astrophysical processes and sources.
At frequencies below $\lesssim $ 1 GHz, the synchrotron radiation from our galaxy dominates, and the thermal bremsstrahlung (free-free) from warm and hot gas in the interstellar medium gives an additional contribution, mostly localized at low galactic latitudes.
Between 30 and 70 GHz, the frequencies covered by the  {\sc Planck-Lfi}, the CMB is the most relevant emission.  Free-free, synchrotron emission and the radiation from interstellar dust  also contribute.

Clearly, any additional signal from DM annihilations/decays, should account only for a (small) fraction of the observed emission. Therefore, a careful evaluation of the astrophysical foreground is mandatory in order to search for DM or to set realistic constraints.
Here we focus on the six radio maps from 22 MHz to 2.3 GHz presented in sec.~\ref{sec:maps}. In principle, we could also include the {\sc Planck-Lfi} maps. However, as it will be clear from sec.~\ref{sec:constraints},  these frequencies only help to slightly improve the constraints for DM masses $\gtrsim$ 100 GeV. Moreover, as discussed before,  several  astrophysical processes are  important at the {\sc Planck-Lfi}  frequencies, which complicate the modeling of the astrophysical foreground. 
For these reasons, we only use {\sc Planck-Lfi} maps to compute conservative bounds in sec.~\ref{sec:constraints}, without assuming any astrophysical model.  We refer the reader to \cite{Egorov:2015eta} for an example of an analysis of {\sc Planck-Lfi} maps with an astrophysical component separation.

Our model of the astrophysical foreground at the 22 MHz -2.3 GHz frequencies  is based on ref.~\cite{Fornengo:2014mna}~\footnote{See also~\cite{Bringmann:2011py,DiBernardo:2012zu,Orlando:2013ysa} for studies of the galactic synchrotron emission.  A model of the diffuse emission based on a principal component analysis has been presented in~\cite{deOliveiraCosta:2008pb}. We do not employ this model in our analysis. We believe that this model would be more appropriate to take into account additional diffuse emissions with a spectral dependence different from that of the astrophysical emission, which is in general not the case for DM (see fig.~\ref{fig:DMTspectradata} and~\ref{fig:DMTspectra}).
}.  The total emission is described as the contribution of  an isotropic component,  a galactic diffuse emission model and a template for DM annihilations.
The galactic  model contains the synchrotron radiation produced by the interactions of  cosmic rays with the galactic magnetic field. Therefore, this component depends on the details of the cosmic ray propagation, the structure of the magnetic field, and the distribution of the cosmic-ray sources.
Here, we are not interested in performing a full scan of the parameter space and determine the uncertainties associated with the modeling of the galactic emission. Rather, we aim at obtaining a satisfactory description of the radio maps with a motivated astrophysical model, and studying the impact of such model on the constraints on exotic DM signals.
Therefore, we consider one of the galactic synchrotron models analyzed in~\cite{Fornengo:2014mna} \footnote{We take model L8a in Table 2.}, keeping in mind that different choices could slightly affect our findings.

The radio sky presents several extended structures, called radio loops. These emissions are thought to be originated by local sources, in particular shells of old supernova remnants. We attempt to trace these extended features through an extra template, obtained from a polarization map at 1420 MHz (see ~\cite{Fornengo:2014mna}).
In summary, our model of the radio sky can be described as:

$$T^{\rm model}(\ell,b)= c_1 + c_2 \, T^{\rm G}(\ell,b) + c_3 \, T^{\rm PI}(\ell,b) +c_4 \, T^{\rm DM}(\ell,b),$$

 where $\ell$ and $b$ are the galactic latitude and longitude. The $T^G$, $T^{PI},$ and $T^{DM}$ denote respectively the galactic diffuse synchrotron template, the template for the radio loops and that one from DM annihilations/decays.
 Each template is normalized by a constant, $c_i,$ which is  determined performing a template fitting analysis of the radio maps.
 Obviously, the coefficient $c_1$ gives the intensity  of the isotropic emission.  Moreover, as a sanity check, we verified that the value of the coefficient $c_2$,  recovered by the fit, is close to one, so that the galactic model  employed in the analysis does not deviate excessively from that one preferred by the data. 

\medskip

The template fitting analysis is performed minimizing the following $\chi^2:$
$$\chi^2=\sum_i \frac{(T^{\rm map}_i-T^{\rm model}_i)^2}{\sigma_i^2},$$
where $i$ runs over the pixels and the variance $\sigma_i$ is obtained combining various sources of errors, namely experimental uncertainties and an empirical estimation of the small scale fluctuations of the map.
The rationale of the latter uncertainty is to account for  the turbulent nature of the magnetic field, and the inhomogeneities of the interstellar medium. These effects introduce fluctuations of the synchrotron intensity at small scales, which are not captured by our modeling of the emission.
Following~\cite{Fornengo:2014mna} ,  we estimate this uncertainty computing, for each pixel of the map, the temperature variance in an angular region of diameter of $\sim15$ degrees centered around the pixel (for more details see sec.~4 of~\cite{Fornengo:2014mna} and see also \cite{Jansson:2012rt} for similar considerations).
It turns out that, in some region of the sky, this quantity dominates over the experimental error of the survey. Thus, the $\chi^2$ will depend on the method that we have adopted to determine the variance (for instance the size of the angular region).
In practice, this astrophysical variance adds to the other sources of uncertainty affecting the astrophysical model.
As stressed before, we are not interested to perform a comprehensive analysis, tackling all the uncertainties discussed so far, which it is actually a very complicated task. Instead, adopting a realistic set-up, we give an indication of how much the {\em conservative} bounds can be ameliorated including astrophysical emissions.

\medskip

Finally we conclude noting that the foreground model described so far has been  constructed in order to reproduce the observations outside the galactic plane. At low galactic latitudes,  a more sophisticated model would be necessary in order, for instance, to account  for bright point sources.
For this reason, we will avoid these regions in our analysis.


\section{Constraints on DM}
\label{sec:constraints}

In this section we present our results in terms of constraints on DM annihilation or decay. Before discussing those, however, we specify the regions of interest that we use in the analysis. 

\begin{figure}[!t]
\begin{center}
\includegraphics[width= 0.8 \textwidth]{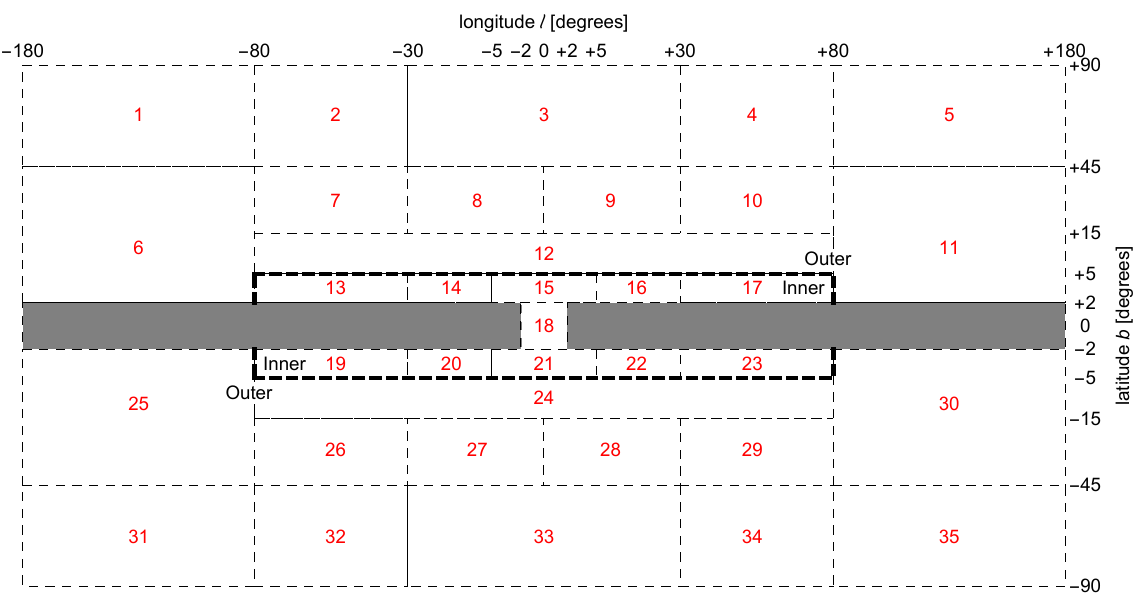}
\caption{\em \small \label{fig:RoIs} {\bfseries Regions of Interest (RoIs)} that we consider. The areas in grey are masked. The thicker dashed lines separate what we define as `Inner Galaxy' and `Outer Galaxy'. We use both for the conservative bounds, we restrict to the Outer Galaxy for the progressive bounds.}
\end{center}
\end{figure}

\subsection{Regions of Interest}
\label{sec:RoIs}

Following~\cite{Cirelli:2015bda}, we divide the whole galactic sky observed by the radio surveys in 35 non-overlapping `regions of interest' (RoI), as depicted in fig.~\ref{fig:RoIs}, masking out the $2^\circ$ around the galactic plane (but retaining a $2^\circ \times 2^\circ$ region around the GC). The regions are designed to be smaller near the GC and wider at high latitude and longitude. For reasons that will be clear later, we distinguish between the Inner Galaxy ($|b| < 15^\circ$, $|\ell| < 80^\circ$, RoI's from 13 to 23) and the Outer Galaxy (the corresponding complement).\footnote{Note that this is slightly different from the analogous distinction used in~\cite{Cirelli:2015bda}.} Some of the radio maps do not cover all the sky. In these cases, we include in the analysis only the RoI presenting a significant amount of pixels.

\subsection{Conservative constraints}
\label{sec:conservative}

\begin{figure}[p]
\begin{center}
\includegraphics[width= 0.48 \textwidth]{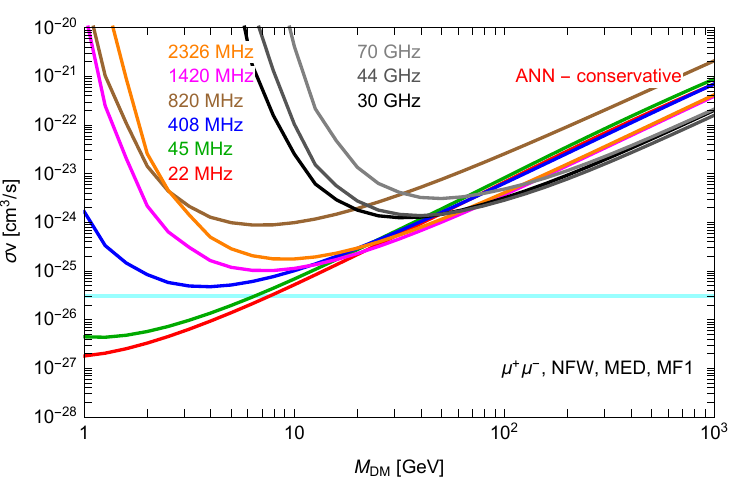} \hfill 
\includegraphics[width= 0.48 \textwidth]{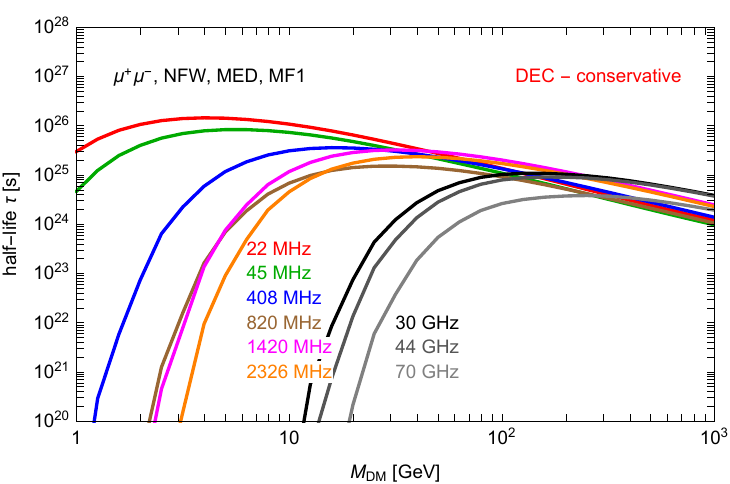} \\
\includegraphics[width= 0.48 \textwidth]{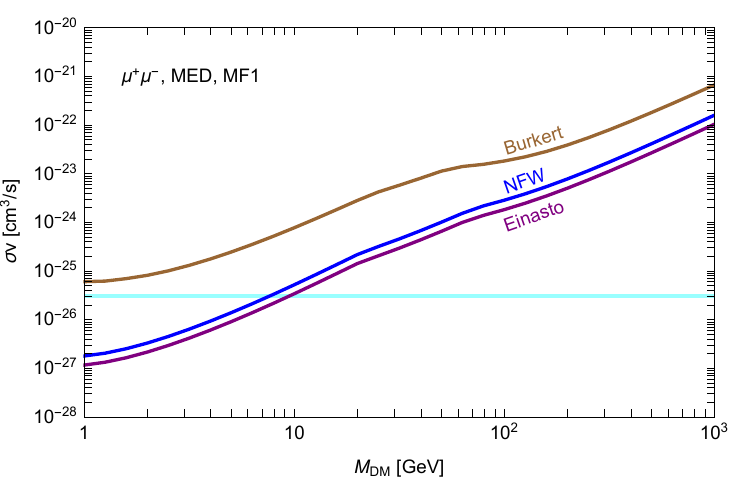}\hfill
\includegraphics[width= 0.48 \textwidth]{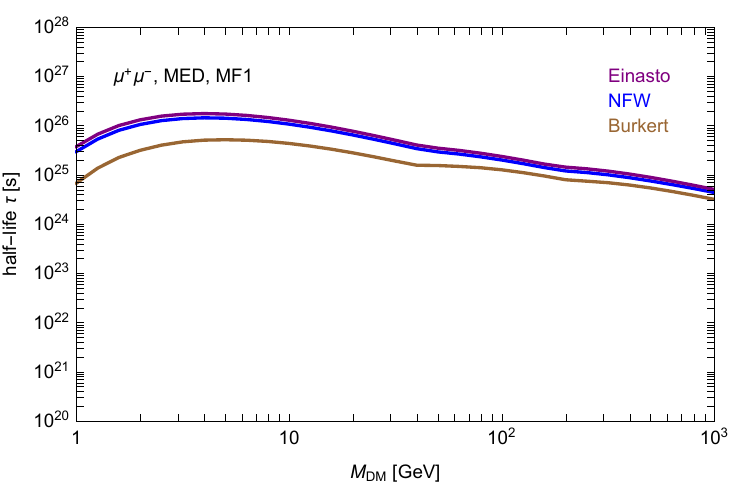}\\
\includegraphics[width= 0.48 \textwidth]{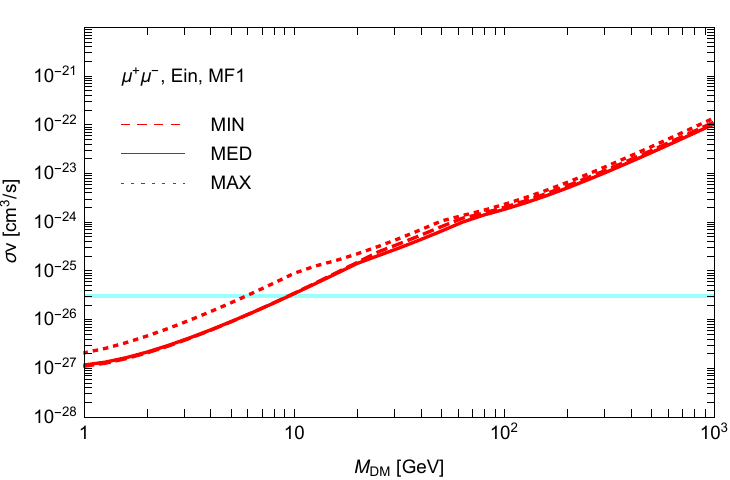}\hfill
\includegraphics[width= 0.48 \textwidth]{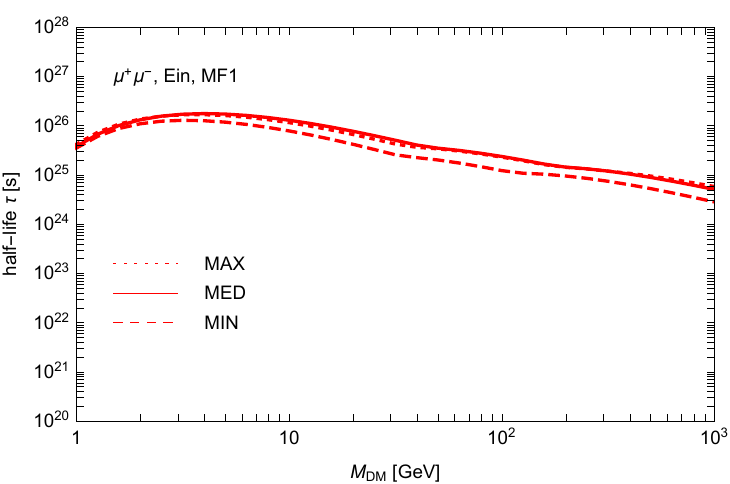}\\
\includegraphics[width= 0.48 \textwidth]{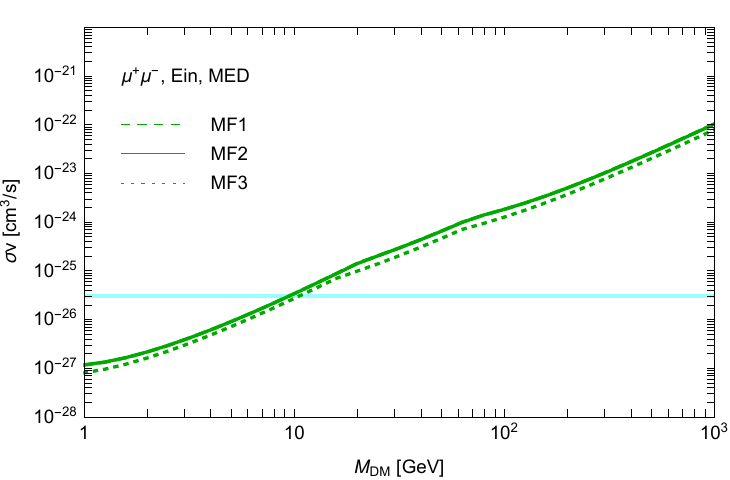}\hfill
\includegraphics[width= 0.48 \textwidth]{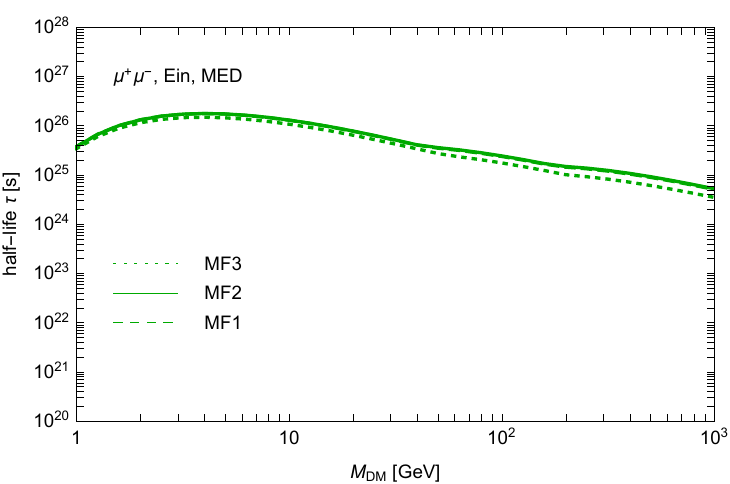}
\caption{\em \small \label{fig:bounds} {\bfseries Dependence of the bounds} on the choice of the observational map (first row), DM profile (second row), galactic $e^\pm$ propagation (third row) and magnetic field (bottom row), for the annihilating DM case (left) and decaying DM case (right).}
\end{center}
\end{figure}

In this section we derive constraints on DM annihilations (and decays) without any assumption on the astrophysical foreground emission: we compute the DM emission for a specific DM model and require it not to exceed the measured radio intensity, in any RoI and for any map. More precisely, for a given annihilation/decay channel, a given DM mass and for a given astrophysical setup (e.g. Ein, {\sc Med}, {\sc Mf1}), we fix a frequency and we compute the integrated emission from DM in each one of the 35 RoI's. Such signal is normalized by the annihilation cross section $\langle \sigma v \rangle$ or the decay rate $\Gamma$. We then compare the signal to the integrated measured intensity in the same RoI and we determine an upper bound to $\langle \sigma v \rangle$ or $\Gamma$ by requiring that the signal does not exceed the measurement plus its error~\footnote{We take the errors on the maps from~\cite{Fornengo:2014mna} and sum all in quadrature. In any case, given the smallness of such errors, the impact on the derived bounds is really limited.}. We then select the RoI which yields the most stringent bound. Finally, we repeat the same procedure for all frequencies and again choose the most stringent. This yields the constraints presented in fig.~\ref{fig:allbounds}, for a specific astrophysical model (Einasto, {\sc Med} and {\sc Mf1}). 

\medskip 

The steps leading to these results and their dependence on different astrophysical setups can be grasped by inspecting fig.~\ref{fig:bounds} and the top row of fig.~\ref{fig:relativeboundsConservative}. In fig.~\ref{fig:bounds} (first row) we show the constraints produced by each individual map: consistently with the discussion in sec.~\ref{sec:formalism}, we see that the low-frequency maps dominate the bound at low masses, while the high-frequency ones are more important at large masses. The global bounds discussed in the following and presented in fig.~\ref{fig:allbounds} are just the envelop of these different curves. In fig.~\ref{fig:bounds} (second row) we show the impact of changing the DM galactic profile: moving from a cuspy profile such as NFW or Einasto to a cored profile such as Burkert relaxes the constraint by a significant amount, up to one or almost two orders of magnitude (depending on the DM mass). In the third and fourth row we show that the impact of changing the propagation models and the magnetic field is instead rather limited. 

Finally, fig.~\ref{fig:relativeboundsConservative} addresses the question of which RoI dominates the constraints. For a specific choice of DM model, a specific astrophysical setup and a given frequency (1420 MHz), we shade each RoI with a color intensity proportional to the strength of the bound that it imposes. We see that, not surprisingly, the regions close to the GC are the most dominant ones. However, it is not the region {\em at} the GC (labelled 18) which wins, and other regions in the inner galaxy are not far behind. This is similar but less pronounced in the case of decay, in which the difference in the constraining power of the RoIs is smaller, consistently with the fact that the signal depends on the first power of the DM density.

\begin{figure}[t]
\begin{center}
\includegraphics[width= 0.48 \textwidth]{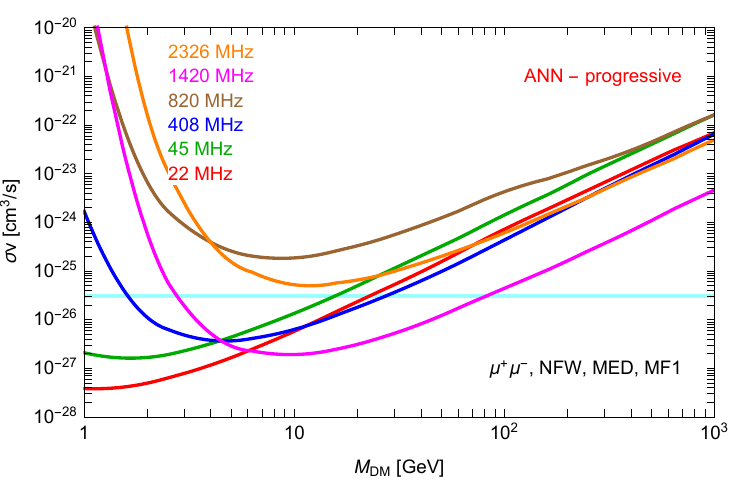}\hfill 
\includegraphics[width= 0.48 \textwidth]{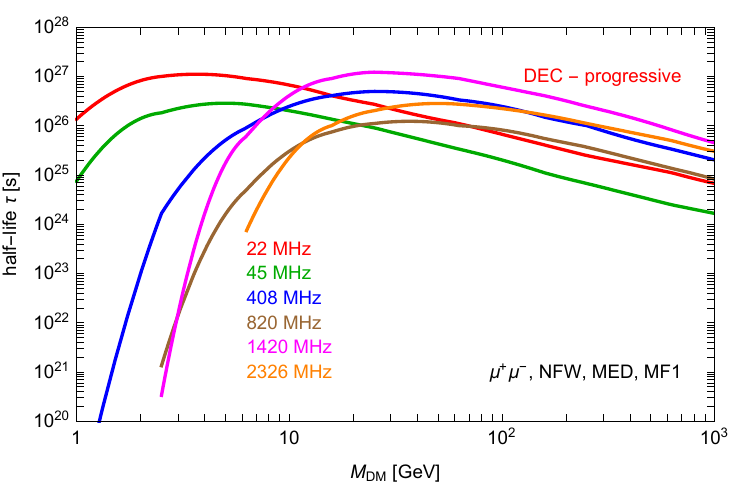} 
\caption{\em \small \label{fig:BoundsProgressiveMapByMap} Bounds from the {\bfseries different maps} in the progressive case.}
\end{center}
\end{figure}

\begin{figure}[!t]
\begin{center}
\includegraphics[width= 0.48 \textwidth]{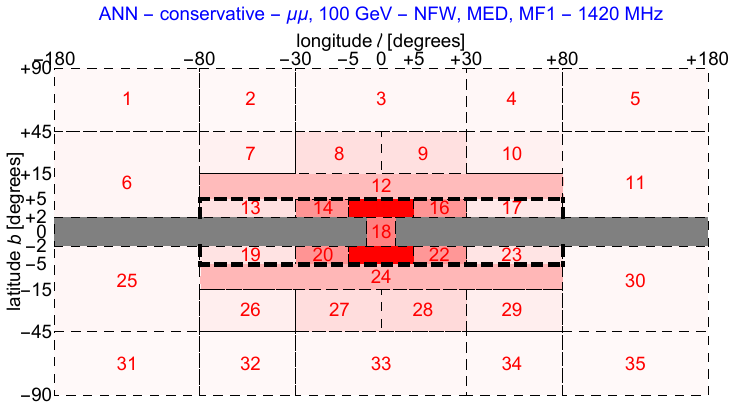}\hfill 
\includegraphics[width= 0.48 \textwidth]{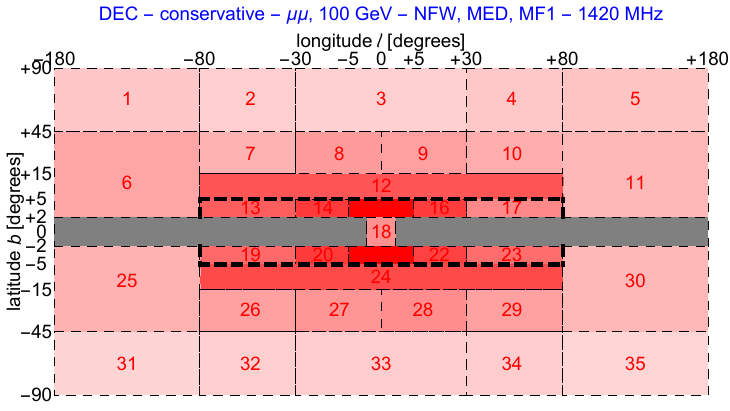} \\[0.5cm]
\includegraphics[width= 0.48 \textwidth]{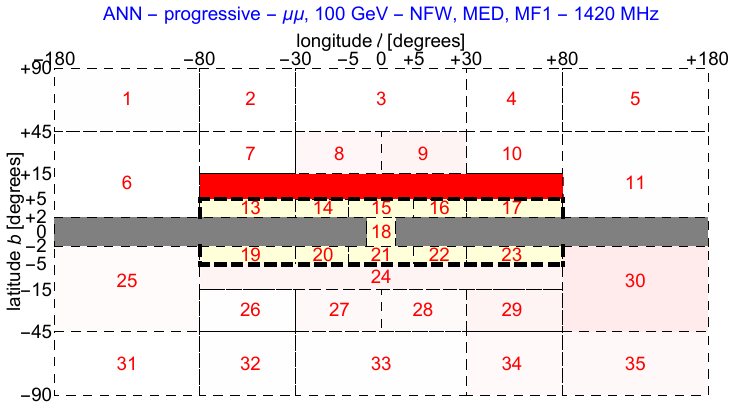}\hfill 
\includegraphics[width= 0.48 \textwidth]{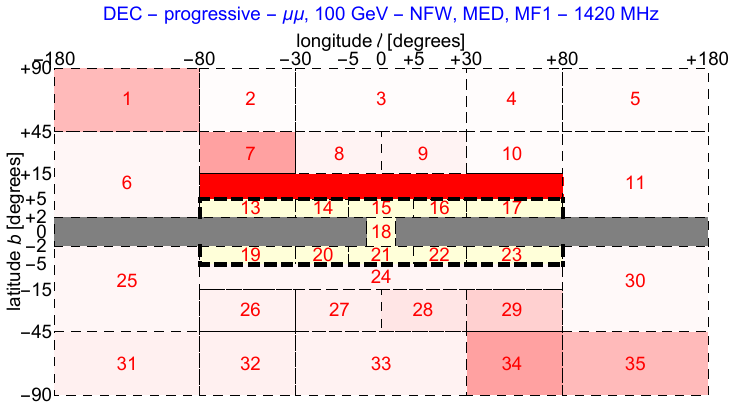} 
\caption{\em \small \label{fig:relativeboundsConservative} {\bfseries Relative strength of the bounds} from the different RoIs (the redder the shading, the more stringent the constraint), for the conservative case (top row) and progressive case (bottom row).}
\end{center}
\end{figure}

\subsection{Progressive constraints}
\label{sec:progressive}

In this section we derive constraints on DM annihilations (and decays) including the astrophysical foregrounds. We consider only the RoIs in the `Outer Galaxy'  in fig.~\ref{fig:RoIs}, in order to avoid low latitude regions in which our foreground modeling is less under control, as discussed in sec.~\ref{sec:background}.
For each of those RoI, we perform a template fitting analysis, as described in~\ref{sec:background}. We remind that we consider four components in the fit: the isotropic emission, a synchrotron emission model from cosmic rays, a template for extended local sources, and a template for DM annihilations.

\begin{figure}[!t]
\begin{center}
\includegraphics[width= 0.49 \textwidth]{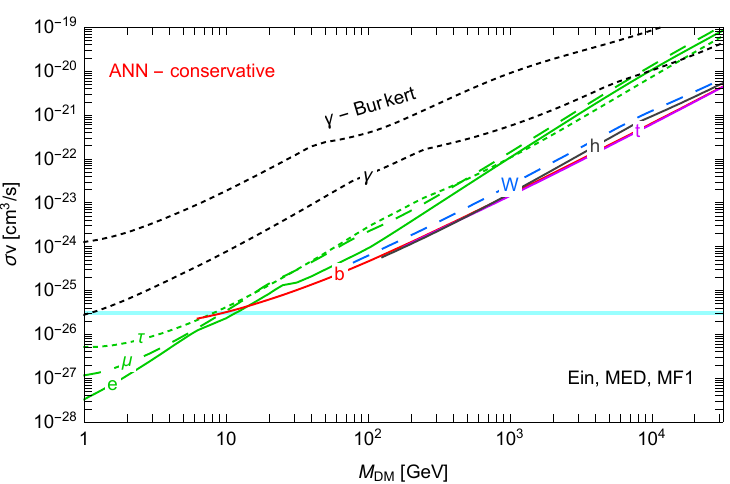} \hfill
\includegraphics[width= 0.49 \textwidth]{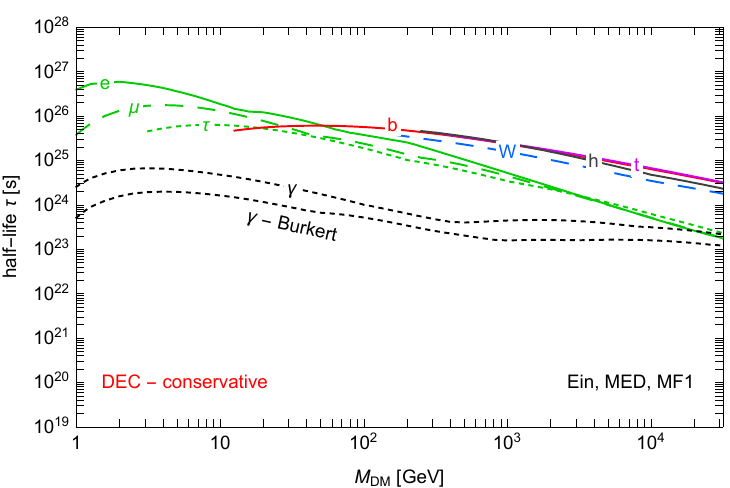} \\

\includegraphics[width= 0.49 \textwidth]{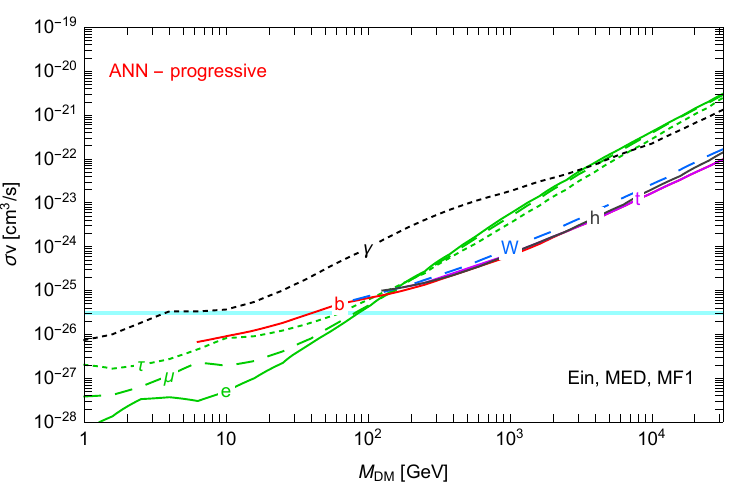} \hfill
\includegraphics[width= 0.49 \textwidth]{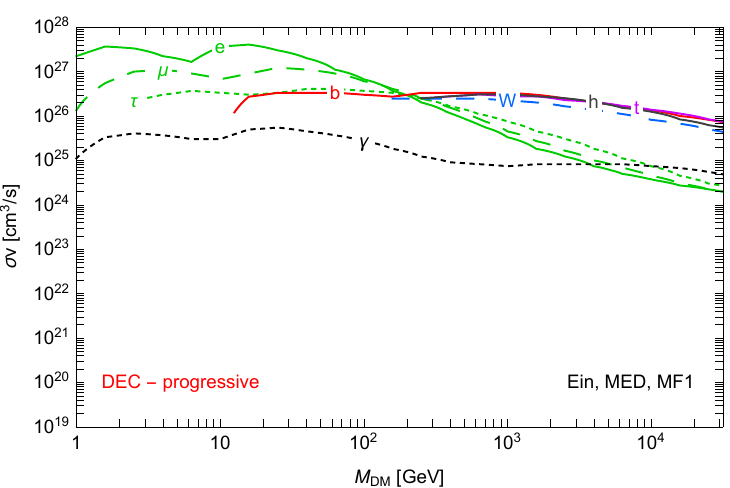} 

\caption{\em \small \label{fig:allbounds} {\bfseries Final bounds} on annihilating DM (left) and decaying DM (right) for all channels. In the top two panels we also show the constraints that apply to the case of a Burkert profile.}
\end{center}
\end{figure}

For each map and for each RoI, we compute the $3 \sigma$ upper limit on the intensity of the DM template (the coefficient $c_4$), profiling over the other parameters of the fit.
Then, for each DM mass, we can select the RoI and radio maps which gives the strongest bound. We present our results in fig.~\ref{fig:BoundsProgressiveMapByMap}, \ref{fig:relativeboundsConservative} and \ref{fig:allbounds}  for a particular choice of DM density profile, propagation setup and magnetic field model.
We find, as in the analysis of the conservative constraints, that the most relevant frequency depends on the DM mass under consideration, with low frequencies more important for light DM. Moreover, as expected, the most important RoIs are those closer to the GC.
In particular, from fig.~\ref{fig:relativeboundsConservative}, it can be noticed that the bounds at 1420 MHz are dominated by one region, i.e. the RoI 12. It happens that, at this frequency, the astrophysical model leaves little room for a DM signal, while in the symmetric RoI, i.e.~number 24, the fit marginally prefers the presence of a DM contribution (which, we believe, has no statistical meaning, due to the large systematic uncertainties plaguing the analysis). This explain why the bound in the RoI. 24 is  much looser than the one from RoI 12.
This trend is not present (or at a much smaller extent) at other frequencies.
 
Finally, we notice that the bounds computed here are more stringent than the conservative ones obtained in sec.~\ref{sec:conservative}. In principle, the former are more realistic, since they take in account the presence of astrophysical foregrounds, which do exist and constitute the bulk of the observed emission.
On the other hand, the {\em progressive} constraints rely on the modeling of the galactic emission, which in turn depends on assumptions on different quantities, e.g. magnetic fields, models of cosmic rays, extended and point-like sources.
A careful assessment of these uncertainties is a difficult task, and it is beyond the scope of this work. Still, we believe that our  {\em progressive} constraints are realistic, especially because they avoid regions at low latitude.

\section{Conclusions}
\label{sec:conclusions}

\vspace{-0.2cm}

In this paper we have reconsidered the constraints on Dark Matter which are imposed by radio and microwave observations of the Galaxy. We have used state-of-the-art semi-analytical tools (discussed in sec.~\ref{sec:formalism}) to compute the DM radio emission and we have compared it with selected observational maps (presented in sec.~\ref{sec:maps}) in a wide range of frequencies from 22 MHz to 70 GHz, including the three {\sc Planck-Lfi} maps. 

Our main results are reported in fig.~\ref{fig:allbounds}, in terms of constraints on the DM annihilation cross-section or decay half-life, respectively, for a variety of primary channels. We have considered {\em conservative} bounds,  computed without assumptions on the foreground astrophysics, and {\em progressive} ones, computed with a modelization of the foreground, which of course reduces the room available for DM leading to tighter bounds. The modelization is anyway limited to regions in which we consider it still reasonably reliable, so that the progressive constraints are believed to be not excessively aggressive. 
We present the results for a benchmark Einasto profile. In case of a cored profile the bounds can be lifted by more than one order of magnitude (for the annihilation case), as discussed in sec.~\ref{sec:conservative}. Other uncertainties have a smaller impact.

\medskip

By confronting with the previous analysis in~\cite{Fornengo:2011iq}, for the cases in which a comparison is possible, we notice that our bounds are a bit looser at small $m_{\rm DM}$, which is in line with what we expected since we are introducing extra low-energy losses. At large $m_{\rm DM}$, our bounds are a bit tighter for the leptonic channels, mostly due to the inclusion of the {\sc Planck-Lfi} maps, not considered in~\cite{Fornengo:2011iq}.

\begin{figure}[t]
\begin{center}
\includegraphics[width= 0.69 \textwidth]{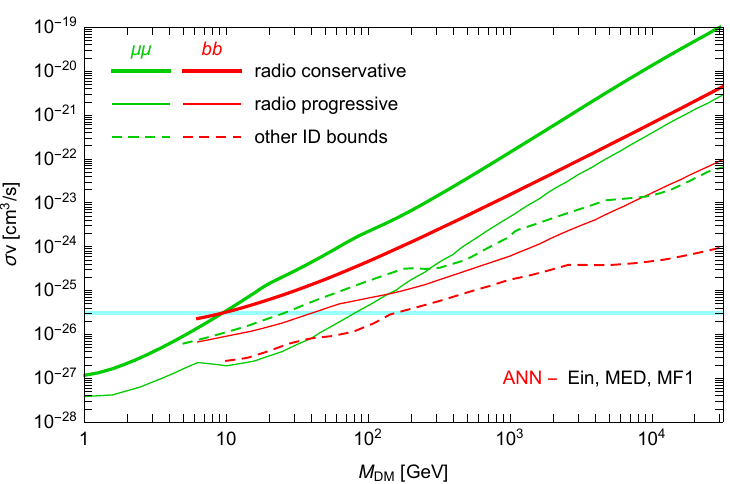}
\caption{\em \small \label{fig:BoundsCompared} {\bfseries Comparison of the radio bounds} derived in this work with the most stringent limits from the combination of other Indirect Detection searches ($\bar p$, $\gamma$-rays and CMB, from \cite{Cirelli:2015gux}).}
\end{center}
\end{figure}

A comparison with other indirect detection bounds on DM (e.g. from high energy $\gamma$-rays or antiprotons, see~\cite{Cirelli:2015gux} for a selection, for the annihilation case) is reported in fig.~\ref{fig:BoundsCompared}. It shows that the conservative radio ones are mostly subdominant, especially for the hadronic and gauge boson channels. The progressive ones can however become dominant for the leptonic channels, especially at small DM masses. 

\medskip

Our work shows therefore the high potential of radio observation as a tool for constraining or discovering a DM contribution in the Galaxy. The most important progress has to come from a better understanding of the astrophysical foreground. 
In this respect, the role of the next generation of radio telescopes {\sc Lofar}~\cite{vanHaarlem:2013dsa,LOFAR} (already active) and {\sc Ska}~\cite{SKA} (whose construction  will start in 2018, with early science expected in 2020) will notably be instrumental.
A reduction of some astrophysical uncertainties on the DM emission, in particular linked to the galactic distribution profile, will also be crucial.

\small
\subsubsection*{Acknowledgments}
We thank C\'eline B\oe hm for some very useful discussions and for the collaboration in the early stages of this work. We also thank Christoph Weniger and Joe Silk for interesting suggestions, plus of course Martti Raidal. For important clarifications on {\sc Planck} maps, we are very grateful to Massimiliano Lattanzi (and Silvia Galli and Jennifer Gaskins). We are also indebted to Oscar Macias for detailed discussions on free-free absorption, and to Andrei Egorov for cross-checks and discussions.  
M.C. acknowledges the hospitality of the Institut d'Astrophysique de Paris ({\sc Iap}) where a part of this work was done.

\medskip

\footnotesize
\noindent Funding and research infrastructure acknowledgements: 
\begin{itemize}
\item[$\ast$] European Research Council ({\sc Erc}) under the EU Seventh Framework Programme (FP7/2007-2013)/{\sc Erc} Starting Grant (agreement n.\ 278234 --- `{\sc NewDark}' project) [work of M.C.],
\item[$\ast$] Centro de Excelencia Severo Ochoa Programme SEV-2012-0249 [work of M.T.],
\item[$\ast$] MINECO through a Severo Ochoa fellowship with the Program SEV-2012-0249 [work of M.T.],
\item[$\ast$] FPA2015-65929-P and Consolider MultiDark CSD2009-00064 [work of M.T.].
\end{itemize}

\bigskip
\appendix

\footnotesize
\begin{multicols}{2}
  
 \end{multicols}

\end{document}